\documentclass{aastex63}
\usepackage{multirow} 
\usepackage{booktabs}

\begin{document}

\title{The Heating and Pulsations of V386 Serpentis after its 2019 Dwarf Nova Outburst}


\author[0000-0003-4373-7777]{Paula Szkody}
\affiliation{Department of Astronomy, University of Washington,
Seattle, WA 98195, USA}

\author[0000-0002-4806-5319]{Patrick Godon}
\affiliation{Department of Astrophysics \& Planetary Science,
Villanova University, Villanova, PA 19085, USA}
\affiliation{Henry A. Rowland Department of Physics \& Astronomy,
The Johns Hopkins University, Baltimore, MD 21218, USA}

\author{Boris T. G\"ansicke}
\affiliation{Department of Physics, University of Warwick, Coventry, CV4 7AL, UK}

\author{Stella Kafka}
\affiliation{AAVSO, 49 Bay State Rd, Cambridge, MA 02138, USA}

\author{Odette F. T. Castillo}
\affiliation{Department of Physics, University of Warwick, Coventry, CV4 7AL, UK}

\author[0000-0002-0656-032X]{Keaton J.~Bell}
\altaffiliation{NSF Astronomy and Astrophysics Postdoctoral Fellow}
\affiliation{Department of Astronomy, University of Washington,
Seattle, WA 98195, USA}

\author{P.~B.~Cho}
\affiliation{Department of Astronomy, University of Texas, Austin, TX 78712, USA }

\author{Edward M. Sion}
\affiliation{Department of Astrophysics \& Planetary Science, Villanova University, Villanova, PA 19085, USA}

\author{Praphull Kumar}
\affiliation{Department of Physics and Astronomy, University of Alabama, Tuscaloosa, AL 35487, USA}

\author{Dean M.~Townsley}
\affiliation{Department of Physics and Astronomy, University of Alabama, Tuscaloosa, AL 35487, USA}

\author{Zach Vanderbosch}
\affiliation{Department of Astronomy, University of Texas, Austin, TX 78712, USA }

\author{Karen I.~Winget}
\affiliation{Department of Astronomy, University of Texas, Austin, TX 78712, USA }

\author{Claire J. Olde Loohuis}
\affiliation{Department of Astronomy, University of Washington, Seattle, WA 98195, USA}

\begin{abstract}
Following the pulsation spectrum of a white dwarf through the heating and cooling involved in a dwarf nova outburst cycle provides a unique view of the changes to convective driving that take place on timescales of months versus millenia for non-accreting white dwarfs. In 2019 January the dwarf nova V386 Ser (one of a small number containing an accreting, pulsating white dwarf), underwent a large amplitude outburst. Hubble Space Telescope ultraviolet spectra were obtained 7 and 13 months after outburst along with optical ground-based photometry during this interval and high-speed photometry at 5.5 and 17 months after outburst. The resulting spectral and pulsational analysis shows a cooling of the white dwarf from 21,020~K to 18,750~K (with a gravity $log(g)=8.1$) between the two UV observations, along with the presence of strong pulsations evident in both UV and optical at a much shorter period after outburst than at quiescence. The pulsation periods consistently lengthened during the year following outburst, in agreement with pulsation theory.  However, it remains to be seen if the behavior at longer times past outburst will mimic the unusual non-monotonic cooling and long periods evident in the similar system GW Lib.

\end{abstract}

\keywords{novae, cataclysmic variables --- stars: white dwarfs --- stars: individual (V386 Ser)}

\section{Introduction}

In the last two decades, almost 20 pulsating white dwarfs have been found to exist in close binaries that have active mass transfer from a late type main sequence star \citep{wvz98,sz12a}. These accreting, pulsating white dwarfs are unique as they undergo dwarf novae outbursts resulting from the the ongoing mass transfer (see review of dwarf nova outbursts in \citet{w95}). During this outburst,
the white dwarf is heated by thousands of degrees \citep{s95,g06}, causing it to move out of the instability strip and the pulsations to cease.
During the subsequent cooling, which takes place on the order of months to years (as opposed to the millenia for the cooling of single white dwarfs), the rapid changes to convective driving can be tracked through the periods of the non-radial pulsations. The period of the most effectively driven mode is expected to scale with the thermal timescale at the base of the convection zone, which is shorter when the outer layers of the white dwarf are heated by the outburst \citep{a06}. Since the non-radial pulsation modes of a white dwarf penetrate deep into the star, these unique systems provide the potential to probe how the accretion of mass and angular momentum affect a star and its subsequent evolution \citep{wk08,fb08,c19,t04}.

While this is a tantalizing exploration, there are several factors that make the corresponding observations difficult. In order to view the pulsations, the mass transfer rate needs to be low and the subsequent disk minimal so that the white dwarf light is a large contribution to the overall observed light. Obtaining an accurate temperature and composition for the white dwarf is only possible in the far ultraviolet, where the white dwarf continuum and absorption lines dominate over the disk emission and lines. This entails using the highly competitive Hubble Space Telescope (HST), and results in much more sporadic and short term coverage than ground observations.  In addition, for low mass transfer rates, the dwarf nova outbursts only recur on decades timescales and are not predictable. 

There is currently only one system, GW Lib, that has been followed extensively before and after its dwarf nova outburst. GW Lib is the first known pulsating accretor and is one of the brightest with V=17 at 
quiescence \citep{wvz98,vz04}. It had a very large (9 mag) amplitude dwarf nova outburst in 2007 followed by five HST/COS observations in 2010, 2011, 2013, 2015, 2017 as well as ground-based optical monitoring throughout this interval \citep{sz12b,sz16,t16,g19,c21}. Those data showed two surprising results. First, the temperature from the UV spectra was not a smooth  cooling transition from the 18,000K determined from the observation 3 yrs after outburst to the 14,700K measured at quiescence \citep{sz02a}. Instead, there was an initial decrease to 16,000K in 2011, and then an increase to the same average higher temperature near 17,000K for the remaining three observations. Second, the pulsation spectrum was not a smooth change from a shorter period after outburst back to the 648, 376 and 236 s observed during quiescence. The pulsation spectrum contained a complex array of short periods (293 s in 2010 and 2011, 275 s in 2013 and 2017 and 370 s in 2015) as well as longer periods at 19 min and 4 hrs that appeared for weeks at a time and then disappeared. Overlapping K2 and HST coverage in 2017 confirmed a correlation of the 4 hr period with a large UV flux increase and the presence of the 275 s pulsation only during that increase. While all the periods appear to relate to different modes of pulsation, the driving mechanism for each mode is not clear and it is obvious that the cooling to quiescence takes more than 10 yrs. 

In order to pursue a better understanding of this unusual behavior, we attempted to observe the behavior of another accreting pulsator (V386 Ser) after it underwent its first dwarf nova outburst in 2019.

\section{Background on V386 Ser}

V386 Ser was discovered as a 19th magnitude cataclysmic variable in the Sloan Digital Sky Survey \citep{sz02b}, showing the broad absorption lines surrounding Balmer emission that is characteristic of low mass transfer systems. Followup photometry by \citet{ww04} identified an orbital period of 80.52 min and pulsations near 607, 345 and 221 s, making it the second known accreting pulsator. A later international campaign was organized, using seven ground observatories over 11 nights. This resulted in a refinement of the main pulsation period to be 609 s and revealed it to be an evenly spaced triplet \citep{m10}, indicating an internal rotation period of 4.8 days. A low resolution HST spectrum with the solar blind channel was obtained during quiescence in 2005 and fit with a quiescent white dwarf temperature of 13,000-14,000K \citep{sz07}. The UV data and simultaneous optical data showed the identical 609 s pulsation with a UV/optical ratio of 6, pointing to a low order pulsation mode.
 V386 Ser underwent its first known dwarf nova outburst on 2019 January 18, and was followed by the AAVSO and other observers. It showed an outburst amplitude of 8 mag followed by several rebrightening events and then a slow decline to optical quiescence (Figure 1).  The basic parameters of V386 Ser are summarized in Table \ref{syspar}.  The distance is derived from the EDR3 Gaia
 parallax of 4.182$\pm$0.276 mas \citep{lin20,lur18}, the E(B-V) from the 3D map by \citet{cap17}, and the white dwarf mass and radius from the gravity we obtain in
 the present work, using the mass-radius relation for a C-O white dwarf \citep{woo95} at
 T $\sim$ 20,000 K.

\begin{deluxetable*}{lcl}[h!]
\tablewidth{0pt}
\tablecaption{Properties of V386 Ser} 
\tablehead{ 
Parameter        & Value        & References        
}
\startdata
Orbital Period              &  80.52 min (4831.2 s)              &   Woudt \& Warner 2004     \\ 
Pulsation periods           & 609 s, 345 s, 221 s                   & Woudt \& Warner 2004, Mukadam et al. 2010  \\
Distance                    & $239^{+17}_{-15}$ pc              &  Gaia EDR3 \\
$E(B-V)$                    & $0.113 \pm 0.04$                  &  Capitanio et al. 2017 \\ 
White Dwarf Mass                     & $0.676_{-0.176}^{+0.216} M_{\odot}$   &  this work        \\ 
White Dwarf Radius                   & $8430_{+2550}^{-2020}$ km          &  this work        \\ 
Log(g)                      & $8.1 \pm 0.36$                     &  this work \\ 
WD Temperature              & 18,750-21,000~K                   & cooling from this work
\enddata 
\tablecomments{The white dwarf mass and radius were obtained from the gravity using the mass-radius
relation for a C-O white dwarf at $T \sim 20,000$~K, as a consequence the $\pm$ uncertainties in the white dwarf
radius are related to the $\mp$ uncertainties in the white dwarf mass.} 
\label{syspar} 
\end{deluxetable*}

\section{Observations}

The optimum first observation of a white dwarf after outburst would take place during the decline from maximum brightness after the dominant accretion disk begins to fade,
to avoid contamination by the latter. However, it needs to be as close to the outburst as possible to get the best measure of the heating that has occurred due to the outburst. An HST mid-cycle proposal was submitted and accepted to obtain early observations and
the first observations with HST were scheduled for 2019 May but failed due to spacecraft jitter. They were re-scheduled as soon as satellite constraints allowed and 4223 s of good exposure time was
obtained on 2019 Aug 15 (7 months after outburst). The Cosmic Origins Spectrograph (COS)
was used in the FUV configuration with the G140L grating
centered at 800 \AA\ , producing a spectrum from 915 \AA\ to 
1945 \AA\  on detector segment A (with detector segment B turned off).  
The data (ldza11010) were collected in TIME-TAG mode and consist of
4 sub-exposures obtained with the spectrum collected in 4 different
positions (FPPOS; slightly shifted relative to one another) 
on the detector, allowing gaps and instrument artifacts to be removed.   
The 4 sub-exposures have lengths of about 
1/5 of the orbital period ($\sim16$ min), with the detailed times listed in Table 2.    

A second set of HST COS spectra were obtained on 2020 Feb 28 with 7384 s of good exposure time and with the COS
instrument set up in exactly the same configuration. 
The data (le8001010) were also collected in TIME-TAG mode, and consist of
6 sub-exposures obtained in the 4 different positions (two positions
were obtained twice).
The 6 sub-exposures are also listed in Table 2 and the times of both HST observations are shown on
the optical light curve of V386 Ser in Figure 1.

All the data were processed through the HST pipeline with CALCOS version 3.3.9.
The resulting summed spectra
are shown in Figure \ref{spectra}, with some basic line identifications in Figure \ref{lineid}, and plots showing the  details of the sub-exposure are 
in Figure \ref{allseg}. 
The flux calibration accuracy of the COS instrument,
now in the {\it third lifetime position} (LP3), is slightly
improved relative to previous locations on the COS FUV
detector and reaches a maximum of 2\% in the wavelengths
considered in this work 
\citep[see the COS instrument report][for details]{deb16}. 

In preparation for the FUV spectral analysis, 
we dereddened the spectra of V386 Ser, using the three-dimensional map of the local interstellar
medium (ISM) extinction by \citet{cap17}\footnote{http://stilism.obspm.fr/} 
to assess the 
reddening. At a distance of nearly 240~pc (from the
Gaia EDR3 parallax) we have $E(B-V)=0.113 \pm 0.04$. 
We computed the extinction law using the analytical expression of 
\citet{fit07}, which we slightly modified to agree with an 
extrapolation of \citet{sav79} in the FUV range
\citep[as suggested by][]{sas02,sel13}.

\begin{deluxetable*}{rrcccr}[h!]  
\tablewidth{0pt}
\tablecaption{HST Observation Log} 
\tablehead{ 
Instrument & Configuration & DATA ID   & Date (UT) & Time (UT) & Exp.time  \\ 
sub-exposure \# & Position \# &       & YYYY Mon DD & hh:mm:ss & sec       \#   
}
\startdata
HST COS/FUV & G140L (800) & LDZA11010 & 2019 Aug 15 & 22:15:49 & 4223    \\ 
             1 &  1       & LDZA11AIQ & 2019 Aug 15 & 22:15:49 & 975    \\  
             2 &  2       & LDZA11AKQ & 2019 Aug 15 & 22:35:05 & 914    \\  
             3 &  3       & LDZA11B1Q & 2019 Aug 15 & 23:42:53 & 1166   \\  
             4 &  4       & LDZA11B9Q & 2019 Aug 16 & 00:05:20 & 1167   \\  
HST COS/FUV & G140L (800) & LE8001010 & 2020 Feb 28 & 19:42:54 & 7384    \\  
             1 &  1       & LE8001A6Q & 2020 Feb 28 & 19:42:54 & 1841   \\  
             2 &  2       & LE8001AGQ & 2020 Feb 28 & 20:14:20 &  310   \\  
             3 &  2       & LE8001AOQ & 2020 Feb 28 & 21:11:28 & 1541   \\  
             4 &  3       & LE8001AQQ & 2020 Feb 28 & 21:38:54 & 1080   \\  
             5 &  3       & LE8001ASQ & 2020 Feb 28 & 22:46:49 &  771   \\  
             6 &  4       & LE8001AUQ & 2020 Feb 28 & 23:01:35 & 1840   \\  
\enddata
\tablecomments{
With the G140L grating center at 800 \AA , 
the detector segment B is turned off.  
The date/time (columns 4 \& 5) refer to the start of the observation.
The exposure time (last column) is the total good exposure time.  
}   
\label{obslog}  
\end{deluxetable*}

Optical observations prior to and during the HST observations were conducted by
AAVSO members to monitor the state of the system (Figure 1). High speed photometry was accomplished at McDonald Observatory on 2019 July 3, 24 and 25 (in the month before the August HST spectra) with a Princeton Instruments ProEM frame-transfer CCD on the 2.1m Otto Struve telescope through a BG40 filter to reduce sky noise, and again on 2020 June 22 (four months after the later HST spectra).  Aperture photometry was performed with the IRAF routine \verb+ccd_hsp+ \citep{2002A&A...389..896K}
while {\sc phot2lc}\footnote{\url{https://github.com/zvanderbosch/phot2lc}} was used to extract light curves using an aperture size that maximized signal-to-noise. 
A journal of observations from McDonald is provided in Table~\ref{mcdonaldlog}.
Observations were also attempted at Apache Point Observatory but were weathered out.

\begin{deluxetable}{ccc}[h!]  
\label{mcdonaldlog} 
\tablewidth{0pt}
\tablecaption{McDonald 2.1-meter Observation Log (ProEM camera, BG40 filter)} 
\tablehead{ 
Date (UT) & Exp. time & No. frames \\ 
YYYY Mon DD & sec   &  \#   
}
\startdata
2019 Jul 03 & 20 & 327 \\
2019 Jul 24 & 10 & 730 \\
2019 Jul 25 & 10 & 1174 \\
2020 Jun 22 & 15 & 747 \\
\enddata
\end{deluxetable}

\section{HST COS FUV Spectral Analysis} 

The suite of codes TLUSTY/SYNSPEC \citep{hub88,hub95} were used
to generate synthetic spectra for high-gravity stellar atmosphere 
white dwarf models. A one-dimensional vertical stellar atmosphere structure is 
first generated with TLUSTY for a given surface gravity ($log(g)$),
effective surface temperature ($T_{\rm eff}$), and surface composition. 
Subsequently, the code SYNSPEC is run, using the output from TLUSTY 
as an input, to solve for the radiation field and generate a synthetic 
stellar spectrum over a given wavelength range between 900~\AA\ and 7500~\AA . 
The code includes the treatment of the quasi-molecular satellite lines
of hydrogen which are often observed as a depression around 1400\AA\ 
in the spectra of white dwarfs at low temperatures and high gravity.
Finally, the code ROTIN is used to reproduce rotational and instrumental
broadening as well as limb darkening. In this manner, 
stellar photospheric spectra covering a wide range of effective temperatures
$T_{\rm eff}$ and surface gravities $Log(g)$ were generated.  

The fitting of the observed spectra with theoretical spectra is carried
out in two distinct steps. 

{\bf (1)} 
In the first step, using the theoretical model spectra we generated
with solar composition
and a {\it standard} projected rotational velocity 
($V_{\rm rot} sin(i)$) 
of 200km/s
(which is common for cataclysmic variables), 
we fit the spectrum for each of the two epochs individually to obtain 
the surface temperature ($T_{\rm eff}$ 
and gravity ($log(g)$) of the white dwarf. This is done in a self-consistent
manner, namely: the gravity obtained for both epochs has to be the same,    
and the best-fit models have to scale to the known Gaia distance.   

Explicitly, 
we generate a grid of solar composition white dwarf models, with a projected
stellar rotational velocity of 200 km/s, an effective surface temperature 
from 17,000~K to 27,000~K in steps of 500~K, and an effective surface
gravity from $log(g)=7.0$ to $log(g)=9.0$ in steps of 0.2,  
a total of $21 \times 11 = 231$ initial models. 
The grid of models is further refined as needed in the area of interest 
(where the best-fit solutions are found in the $log(g)$ vs. $T_{\rm wd}$  
parameter space) 
by generating additional models in steps of 250~K in temperature
and steps of 0.1 in $log(g)$. 
For each white dwarf temperature and gravity we derive the white dwarf radius $R_{\rm wd}$ 
and mass $M_{\rm wd}$ by using the (non-zero temperature) C-O white dwarf 
mass-radius relation from \citet{woo95}. For each model fit, using the 
white dwarf radius and scaling the theoretical flux to the observed flux, 
we derive a distance. As each observed COS spectrum is fitted to the
models in the grid, the fitting yields a reduced $\chi^2_{\nu}$ ($\chi^2$ per 
degree of freedom $\nu$) value and a distance $d$ for all the (grid) points
in the parameter space ($log(g),T_{\rm wd}$). 
We then find the model for which $\chi^2$ is minimum
($\chi^2_{\rm MIN}$) amongst all the models that scale to
the distance $d$.  

Note that before the fitting, we mask the Ly$\alpha$ region and the 
O\,{\sc i} (1300 \AA ) region, both contaminated by airglow, 
as well as the C\,{\sc iv} (1550 \AA ) broad emission line
since it does not originate in the WD photosphere. 
Since we wish to derive the temperature and gravity, we fit
the Ly$\alpha$ wings and the continuum slope of the spectra
by masking prominent absorption lines (an accurate fit to the
absorption lines is carried out in the second step).

The value of $\chi^2$ is subject to noise, which is inherited
from the noise of the data. As a consequence there is an 
uncertainty in $\chi^2$, which translates into uncertainties
in the derived value of $log(g)$ and $T_{\rm wd}$ - 
{\it the statistical errors} as opposed to systematic errors. For a number of parameters
$p$, the uncertainty on the derived parameters values
is obtained for $\chi^2$ within the range 
$\chi^2_{\rm MIN}$ and $\chi^2_{\rm MIN} + \chi^2_p (\alpha)$, 
where $\chi^2_p(\alpha)$ is for a significance $\alpha$,
or equivalently for a confidence $C=1-\alpha$ 
\citep[see ][]{lam76,avn76}.

 (2) 
In a second step, 
once a best fit $(T_{\rm eff}, log(g))$ is found for each epoch, 
we vary the abundances of specific elements 
(e.g. Si, C,..) one by one and vary the projected stellar rotational 
velocity, $V_{\rm rot} sin(i)$, in the models 
until the absorption lines for each element are fitted.

\subsection{White dwarf temperature and gravity}

\subsubsection{The August 2019 COS Spectrum Analysis} 

We first carried out a spectral analysis of the 2019 Aug COS spectrum, 
dereddening the spectrum with $E(B-V)=0.113$, using our coarser grid of
models in steps of 500~K in temperature and 0.2 in log(g) and masking the 
emission and absorption lines. The preliminary
results indicated that the best fit solutions scaling to the correct distance 
were in the range $7.6 < log(g) < 8.6$ and 18,000~K $< T_{\rm wd} <$ 24,000~K. 
We then continued with models in this region of the parameter space using our
finer grid of models in steps of 250~K in temperature and 0.1 in log(g). 
The results are then obtained in the log(g) vs. $T_{\rm wd}$ two-dimensional 
parameter space: for each (model) grid point in the two-dimensional
parameter space we obtain a 
reduced $\chi^2_{\nu}$ value and a distance $d$. 
The distance $d$ is first treated as a free parameter. 
Then, the constraint $d=239$~pc is used to reduce the problem
from a two-dimensional problem into a one-dimensional problem:
to find the  model with the least $\chi^2_{\nu}$  amongst all 
the models scaling to $d=239$~pc (which form a line in the
two-dimensional parameter space). 

The smallest value $\chi^2_{\nu {\rm MIN}}$ in the two-dimensional
parameter space is $\sim0.446$, significantly
smaller than 1. This can happen if the errors in the observed data 
(i.e. errors in the flux $F_{\lambda}$, in erg/s/cm$^2$/\AA ) are large 
(or overestimated). 
Also, in the second step, where we model the absorption lines 
and significantly reduce the masking of the spectrum, we obtain
a $\chi^2_{\nu {\rm MIN}}$ closer to one. 

We summarize this first set of results in Figure \ref{aug19achi2}, 
where we draw a map of the $\chi^2_{\nu}$ values in the 
$log(g)$ vs. $T_{\rm wd}$ parameter space. Each model forms a rectangle
of size 0.1 (in $g$) $\times$ 250 (K) and is colored in grey according to 
its $\chi^2_{\nu}$ value: smaller $\chi^2_{\nu}$ is darker. For convenience and 
clarity only the region of interest is shown, forming a (dark gray) diagonal, 
and the remaining models (with a higher $\chi^2_{\nu}$ value) have been left in white). 
The three white and blue dashed diagonal lines represent the scaling to the Gaia distance
$d=239^{+17}_{-15}$~pc as indicated in the lower right of the panel. 
Superposed to this is a contour map (in yellow) of the $\chi^2_{\nu}$ 
values extrapolated from the models (rectangles) which we use to find the 
least $\chi^2_{\nu}$ along the distance lines.
Along the d=239~pc curve (middle dashed white blue line), the 
least chi square model is obtained where the middle red dot is located, 
and yields
$T_{\rm wd}= 21,230$~K with 
$log(g)=8.08$. This model has 
$\chi^2_{\nu}=0.475$ and the spectral fit is shown in Figure \ref{T21230}. 

\paragraph{The Distance} 
In order to assess how the error in the Gaia distance
$d=239_{-15}^{+17}$~pc propagates, in Figure \ref{aug19achi2} 
we also draw the lines for which the models scale to a 
distance of 224~pc (right dashed blue line) and 256~pc 
(left dashed blue line). 
The intersection of these lines 
with the least chi square diagonal results in 
errors rounded up to $\pm 0.2$ in $log(g)$ and 
$\pm 500$~K in $T_{\rm wd}$. 

\paragraph{The Grid of Models} 
Since the models were generated in steps of 250~K in temperature
and 0.1 in $log(g)$, we estimate that the solution
is accurate to within about half these values: 
$\sim 0.05$ in $log(g)$ and $\sim 125$~K in $T_{\rm wd}$. 

\paragraph{The Reddening}
To assess how the error in the reddening affects the 
results, we dereddened the spectrum within the limits of the error bars on $E(B-V)$
i.e.
$E(B-V)=0.073$ and $E(B-V)=0.153$ (Table \ref{syspar}) and carried  
out the same spectral analysis for these two values with the fine grid of models. 
The results ( Figure \ref{aug19_reddening}) show that the reddening uncertainty of 
$\pm 0.04$ yields an error of $^{-450}_{+395}$~K in temperature
and $^{-0.29}_{+0.27}$ in $log(g)$, where the lower temperature
is for the larger dereddening. While the larger dereddening increases
the temperature of the gray {\it diagonal} by about $\sim 1000$~K (the spectrum becomes bluer), 
it also increases the flux by a factor of 2 compared to the smaller
dereddening. The spectrum dereddened with $E(B-V)=0.153$ 
has a continuum flux level $\sim$40\% larger than when dereddened 
with $E(B-V=0.113$, itself having a flux $\sim$40\% larger than 
dereddened with $E(B-V)=0.073$. 
As a consequence the solutions that scale to this larger
flux have a larger radius, and therefore lower gravity. Since the 
best-fit solutions (gray diagonal) have a decreasing temperature
with decreasing gravity, the overall solution becomes colder for 
the larger dereddening value $E(B-V)=0.153$. This is counter to the 
simple assumption that a bluer spectrum is hotter, as the simple assumption
doesn't take into account the distance and radius (i.e. gravity) of the white dwarf. 
This is explicitly visualized in Figure \ref{aug19_reddening}. 

\paragraph{The Second Component} 
While the center of the Ly$\alpha$ absorption profile is affected
by airglow emission, the bottom of Ly$\alpha$ (in the region
where it flattens, see Figure \ref{spectra}) is not at zero. 
 It has a flux
about 10 times smaller than the continuum flux level on both sides of
Ly$\alpha$, $\sim 1.35 \times 10^{-15}$erg/s/cm$^2$/\AA\ 
(in the dereddened spectrum). 
This could be due to an elevated white dwarf temperature (above a certain temperature the bottom of Ly$\alpha$ doesn't go to zero) or to the 
presence of a (hotter) second component. Indeed, a second component is often
observed in the spectra of CV white dwarfs at low mass accretion rate,
and while its origin is still a matter of debate, it is 
{\it customary} to model such a component as a flat continuum
\citep[e.g.][]{pal17}. The exact nature of the second component is
unknown, but it is suspected to be either the boundary layer or the inner disk.
We therefore carried out a spectral
fit assuming different values for the flat 2nd component. 
We found that the addition of a second component only slightly improved
the fit in the $\chi^2$ sense: a second component of the order of 
$5 \times 10^{-16}$ yields $\chi^2_{\nu} = 0.46$ (Figure \ref{T21000}), a decrease
of only 3\% in the chi square (no 2nd component had $\chi^2_{\nu}=0.475$).  
The addition of such a second component decreases the temperature
by about $\sim 200$~K and increases the gravity by 0.006 in $log(g)$, 
i.e. $T_{\rm wd}=21,020$~K with $log(g)=8.087$. 
The decrease in $\chi^2_{\nu}$ is rather small, and the changes in 
temperature and gravity are much smaller than due to the errors in 
the distance and reddening. We adopt this solution as
the final result. 

\paragraph{The Statistical Error}

If the problem could be summarized as finding the least $\chi^2_{\nu}$ model 
in the two dimensional parameter space ($log(g),T_{\rm wd}$), 
then the parameter $p$ (see beginning of section 4) would take the value $p=2$.
However, the distance $d=239$~pc enters a constraint and reduces the problem to 
a one-dimensional problem: 
namely, to find the smallest $\chi^2$ value
along the $d=239$~pc line in the ($log(g),T_{\rm wd}$) 
parameter space, rather than in entire two-dimensional (log(g),$T_{\rm wd}$) 
space. 
Therefore, for the statistical error we chose $\chi^2_p(\alpha)$ 
with $\alpha=0.01$ (99\% confidence) and $p=1$: 
for a one parameter problem. 
From \citet{lam76}, we have 
$\chi^2_1(0.01)=6.63$, and the spectral fit has 5453 degrees
of freedom (after masking), giving a value of $\sim 0.0012$ 
for the reduced value of $\chi^2_p$. We find that this produces
an error in temperature of about $\pm$100~K and $\pm 0.02$ in 
$log(g)$, much smaller than all the other errors.

\begin{deluxetable}{lllllcc}[h!]  
\tablewidth{0pt}
\tablecaption{Results for $T_{\rm wd}$ and $Log(g)$ with Error Estimates  
} 
\tablehead{ 
\multicolumn{1}{c|}{         } & \multicolumn{2}{c|}{V386 Ser} & \multicolumn{2}{c}{V386 Ser} & \multicolumn{2}{c}{} \\ 
\multicolumn{1}{c|}{         } & \multicolumn{2}{c|}{COS Aug 19} & \multicolumn{2}{c}{COS Feb 20} & \multicolumn{2}{c}{} \\
\multicolumn{1}{c|}{ } & \multicolumn{1}{c}{$T_{\rm wd}$} & \multicolumn{1}{c|}{$Log(g)$} & \colhead{$T_{\rm wd}$} & \multicolumn{1}{c}{$Log(g)$} & \colhead{} & \multicolumn{1}{c}{} \\\cmidrule{1-7}   
\multicolumn{1}{c|}{Final Best-fit}  & 21,020    & 8.09      & 18,750   & 8.10      &          &           \\ \cmidrule{1-7} 
\multicolumn{1}{c}{}                           & \multicolumn{6}{c}{ Errors  }   \\\cmidrule{1-7}  
\multicolumn{1}{c|}{Source of Errors} & \multicolumn{1}{c}{$\Delta T_{\rm wd}$} & \multicolumn{1}{c|}{$\Delta Log(g)$} & \colhead{$\Delta T_{\rm wd}$} & \multicolumn{1}{c}{$\Delta Log(g)$} & \colhead{} & \multicolumn{1}{c}{}    
}
\startdata
distance $d$         & $\pm 535$ & $\pm 0.2$   & $\pm 535$ & $\pm 0.2$  &   &  \\  
$E(B-V)$             & $\pm 425$ & $\pm 0.3$   & $\pm 425$ & $\pm 0.3$  &   &  \\  
instrument           & $\pm 100$ & $\pm 0.02$  & $\pm 100$ & $\pm 0.02$ &   &  \\  
grid step size       & $\pm 125$ & $\pm 0.05$  & $\pm 125$ & $\pm 0.05$ &   &  \\  
statistical $\chi^2$ & $\pm 100$ & $\pm 0.02$ & $\pm 100$ & $\pm 0.02$&   &  \\  
\hline
Final Results        & $21,020 \pm 710$ & $8.1 \pm 0.36$ & $18,750 \pm 710$ & $8.1 \pm 0.36$ & & \\
\enddata
\tablecomments{The temperature is in Kelvin and the gravity in cgs. 
This value of $log(g)=8.1$ corresponds to a WD mass of $0.676M_{\odot}$ 
at a temperature of $\sim 20,000$~K. The final best-fits are those 
including a flat second component. 
} 
\label{twdlogg} 
\end{deluxetable}

\paragraph{Instrumental Error} 
The flux calibration accuracy of the COS instrument reaches 
about 2\% \citep[][with COS now in LP3]{deb16}, 20 times 
smaller than the change in flux due to reddening errors. 
This 2\% change in flux corresponds to $\sim$100~K in temperature
(at 20,000~K), the same order of magnitude as the statistical error. 

All the errors are recapitulated in Table \ref{twdlogg}. 
The errors, summed in quadrature, are 
$\pm 710$~K in temperature and $\pm 0.36$ in $log(g)$, 
 such that the final result is
$T_{\rm wd}= 21,020 \pm 710$~K with $log(g)=8.1 \pm 0.36$, where
most of the uncertainty in these values is due to uncertainties
in the reddening and distance. 

\subsubsection{The February 2020 COS Spectrum Analysis} 

For the Feb 2020 HST COS spectrum of V368 Ser, we carried out
the same modeling, first with the original spectrum (i.e. without removing
a constant flux). We obtained a temperature roughly 2000~K lower ($T_{\rm wd}=19,250$~K) 
 than for the 2019 Aug spectrum, while the gravity is nearly 
the same (log(g)=8.11 (2020 Feb) vs. 8.08 (2019 Aug))
and the errors add up to the same $\pm 0.36$ in $log(g)$ 
and $\sim \pm 710$~K in temperature. 
For this model $\chi^2_{\nu}=0.50297$. 

We next subtracted a constant flux from the observed spectrum, using
a constant flux of up to $1 \times 10^{-15}$erg/s/cm$^2$/\AA , 
in steps of $1 \times 10^{-16}$erg/s/cm$^2$/\AA , 
and carried out the same spectral modeling. 
The lowest $\chi^2_{\nu}$ was obtained for a 
subtracted flux of $6 \times 10^{-16}$erg/s/cm$^2$/\AA . 
The resulting temperature was $T_{\rm wd}= 18,750$~K with
$log(g)=8.1$ and $\chi^2_{\nu}=0.420$. The $\chi^2_{\nu}$ is reduced
by 16\%. We adopt
this as our best fit final solution for February and show the 
model fit in Figure \ref{T18.75} and the parameters 
in Table \ref{twdlogg}. 

We note that in both the 2019 Aug  and 2020 Feb spectra there is no 
sign of hydrogen quasi-molecular absorption, usually seen near 1400~\AA\ 
at low temperature and high gravity. In order to improve the fit, the
hydrogen quasi-molecular absorption option was turned off in TLUSTY/SYNSPEC
when computing the fine grid of models.

\subsection{White Dwarf Photospheric Abundances and Rotational Velocity}

Both the 2019 Aug (LDZA11010) and 2020 Feb (LE8001010) spectra have
a co-added (good) exposure time of the order of the binary period 
and, therefore, are affected by the orbital motion of the 
white dwarf. In order to derive the white dwarf stellar rotational velocity from the 
absorption lines, we need short-exposure time spectra 
which are not broadened by the orbital motion
of the white dwarf during the duration of the exposure.  
However, because of their short time, the sub-exposures 
have low S/N
that renders the modeling of the lines practically impossible. 
In Figure \ref{allseg}, we display a spectral region with all the 
sub-exposures of both the Aug 2019 and Feb 2020 spectra.
The shortest exposure time is 310 s, which represents only 
a small fraction of the binary period during which the white dwarf
only moved $22^{\circ}$ in its orbit. However, this spectrum
is extremely noisy and cannot be used to derive reliable
abundances and rotational velocity. The total exposure time
of the spectra is not long enough to allow for co-adding
phase-resolved exposures with good S/N. 

Therefore, we used the co-added spectra as in the previous subsection. 
To derive the white dwarf photospheric abundances and rotational velocity, we chose the best fit model temperature and gravity for each spectrum. 
For the 2019 Aug COS spectrum we have $T=21,000$ K, and for the 2020 Feb 
COS spectrum we have $T=18,750$ K, both with $log(g)=8.1$. 
For both models we subtracted a constant flux level of
$6 \times 10^{-16}$ erg/s/cm$^2$/\AA , since both agree with
that subtracted flux for the 2nd component.   

We started with the (co-added) 2019 Aug spectrum and varied
the abundances of C, N, Si, and S, as well as the broadening
velocity. Except for the strong carbon line at 1140~\AA , we found 
that the best-fit was obtained for a solar carbon abundance
with a broadening velocity of $300 \pm 50$~km/s. The C\,{\sc i} 
(1140) absorption line is much smaller in the observed spectrum
than in the model, corresponding to a much lower abundance
of [C]=0.2. 
For the silicon, we found an overall good agreement with 
an abundance of 0.5 solar and the same broadening velocity. A few lines
seem to disagree: the Si multiplet near 1110~\AA\ is better fitted
with solar abundance, while at 1150~\AA , 1265~\AA , 
the silicon better agrees with a low abundance
of $\sim 0.1 - 0.3$ solar. The only discernible nitrogen line (1135) 
and sulphur line (1125) agree both with solar abundance and 
$V_{\rm rot} sin(i)=300$~km/s. It is possible that the C and Si abundance discrepancies
are due to the poor S/N in the short wavelength of the COS 
detector. Near 1250~\AA\ and 1325~\AA\ the continuum does not 
align with the model and could also explain some small discrepancy
there. We present a model with solar abundances, except for
[Si]=0.5, and a broadening velocity of 300~km/s in Figure \ref{aug19ab}. 
For clarity, we only show the spectral region between 1100~\AA\ and 
1450~\AA . 
In the upper panel two short wavelength regions
are not well fitted: 
(i) the C\,{\sc i} (1140) absorption line is much shallower in the 
observed spectrum than in the model, corresponding to an abundance
of [C]=0.2; (ii) the Si multiplet near 1110~\AA\ is better fitted
with solar abundance. 
It is possible that C and Si abundances discrepancy
is due to the poor S/N in the short wavelength of the COS 
detector. 
At $\sim$1250~\AA\ (middle panel) 
the silicon lines better agree with a low abundance
of $\sim 0.1 - 0.3$ solar, however, 
the continuum there doesn't 
align with the model and could explain the some of the discrepancy. 
At $\sim$1280~\AA\ and $\sim$1323~\AA\ 
the carbon lines agree with a subsolar abundance. 
The only discernible nitrogen ($\sim$1135) 
and sulphur ($\sim$1125) lines (see Fig.\ref{lineid}) 
both agree with solar abundance and 
$V=300$~km/s. 
The geocoronal emission region (Ly$\alpha$ and O\,{\sc i} 
$\sim 1300$~\AA ) were masked and are marked in blue.

We then checked by fitting the third exposure of the 2019 Aug data,
obtaining a velocity of 200~km/s and possibly a slightly higher
abundance of carbon based on the C\,{\sc ii} (1175) multiplet. 
It is highly likely that this exposure is still affected by broadening
due to the white dwarf motion during the $\sim 1000$s exposure time, but
not as much as the co-added spectrum with a broadening velocity
of 300~km/s. 

The 2020 Feb  co-added spectrum agrees within the error bars
with the 2019 Aug co-added spectrum, with a broadening velocity 
of about $250 \pm 50$~km/s, solar carbon abundance and subsolar
silicon abundance. However, the Si\,{\sc iii} (1110) multiplet
seems to be solar. Here too, the C\,{\sc i} (1140) line is much
more pronounced in the model than in the spectrum, as well as 
the C\,{\sc i} (1130). As for the 2020 Aug spectrum, the continuum
near the silicon 1250 feature seems to be lower in the model than
in the observed spectrum. Both the C\,{\sc i} and C\,{\sc ii} 
(between 1320 and 1340~\AA ) lines as well as the Si\,{\sc ii}
doublet (near 1530~\AA ) are well fitted at solar abundances 
with a broadening velocity of 200~km/s. 
We did not attempt to fit the 310s (2nd) exposure of the 2020 Feb 
spectrum, as it is far too noisy. 

Velocities of a few hundred km s$^{-1}$ are typical for the fits to the UV spectra
of dwarf novae. The highest resolved spectra of UV absorption lines from $HST$ are
from U Gem \citep{s94}, providing a rotation velocity of 50-100 km s$^{-1}$. The even pulsation
splitting of the 609 s mode of V386 Ser implies a rotation period of 4.8$\pm$0.6 days, or an internal rotation
velocity of $\leq$ 1 km s$^{-1}$ \citep{m10}. Thus, while it appears that differential rotation between the atmosphere and interior of V386 Ser is present, much more data would be needed to obtain a definitive value for the atmospheric rotation.

\section{Cooling Curve}

The cooling curves of dwarf novae that have been measured generally show a smooth transition back to a quiescent temperature e.g U Gem and WZ Sge \citep{g06,g17}, but the timescales have been relatively short due to frequent outbursts or lack of extended data. As described in the introduction, GW Lib has the longest
span of measurements following its outburst and showed that the cooling is not monotonic.
However, the amplitude of its outburst was the largest known and the first measurements of the white dwarf temperature did not take place with $HST$ until 3 yrs after the outburst. The temperatures versus time of GW Lib and V386 Ser are compared in Figure 11. While the quiescent temperatures and the outburst amplitudes are similar, it is too early to tell if V386 Ser will follow the unusual behavior of GW Lib or continue on a normal cooling sequence.

\section{White Dwarf Pulsations and Other Variability}

Having the HST data taken in Time-tag mode allows the creation of light curves with any 
desired time bins. The lightcurves were extracted from the Aug and Feb spectra in the wavelength range between $916-1887\,\AA$. During the extraction process, the geocoronal airglow emission lines of Ly$\alpha$ and O\,{\sc i} were masked out in the range $1207.94-1223.98\,\AA$ and $1296.43-1312.28\,\AA$ respectively. Finally, the data were binned to 5s resolution. A discrete Fourier transform was used to create power spectra to reveal significant periods. The 2019 August 15 data reveal a strong period at 104.284$\pm$0.051 s, while the 2020 Feb 28 data show periods at 174.561$\pm$0.076 and 187.604$\pm$0.090 s. The power spectra are shown in Figure 12. 

We also detect pulsation signals in the high-speed optical photometry from McDonald Observatory. To acquire realistic uncertainties based on the residuals of our fits of pulsation signals to the nightly time series, we detrend each light curve with a 2nd-order Savitzky-Golay filter of width 30 minutes with the W\={o}tan package \citep{wotan}.  This is to remove low-frequency variability from the binary system or variations in weather conditions. The nightly Lomb-Scargle periodograms displayed in Figure~\ref{fig:mcdonald} of both the original and detrended light curves are not significantly different in the frequency range of pulsation signals. We obtain a non-linear least-squares fit of a sinusoid to each detrended light curve with \verb+lmfit+ \citep{lmfit}\footnote{Via the {\sc Pyriod} package: \url{github.com/keatonb/Pyriod}} to obtain the period and amplitude measurements listed in Table~\ref{tab:mcdonaldpers}. The three light curves obtained in the month prior to the 2019 August HST observations exhibit the same periodicity, with a weighted mean period of 104.06$\pm$0.06\,s.  This agrees within $3\sigma$ with the period measured from the HST data. The UV/optical amplitude ratio is $\sim$ 4.5, similar to that at quiescence \citep{sz07} and indicative of a mode with a low spherical degree \citep[$\ell=1\ \mathrm{or}\ 2$;][]{Kepler2000}.

The 2020 June light curve from McDonald does not show the same periodicities as measured from the second set of HST observations obtained roughly four months prior. The 190.0$\pm$0.3\,s optical period is longer than the lower-amplitude UV periodicity by $>7\sigma$; however, this signal could still correspond to the same pulsation mode that has shifted in intrinsic frequency.  We used the Modules for Experiments in Stellar Astrophysics (MESA) and the asteroseismology software Gyre to explore how much pulsational eigenfrequencies can change due to the stellar structure responding to the dwarf nova outburst.  We evolved a 0.93\,$M_\odot$ white dwarf from a 6\,$M_\odot$ main sequence star, then through cooling \citep{Timmesetal2018}, and finally subject to long-term accretion encompassing tens of classical nova outbursts.  Near a midpoint between outbursts, the resulting WD, with a surface temperature of about 15,000~K, was then subjected to a sequence of a few dwarf novae phases with a 30-year recurrence time.  Gyre version 5.2 was then used to perform an adiabatic evaluation of the g-mode frequencies assuming no rotation for structures just before and for some time after a dwarf nova outburst.  Figure~\ref{fig:mesa} shows that the eigenvalue pulsation frequencies all increase rapidly by a small amount in response to the nova outburst, and then slowly drift back toward their quiescent values over many months.  This supports the idea that the same pulsation mode in V386 Ser might have a period 1\% longer in June 2020 than it had in February.

Overall, we observe that pulsation modes with increasingly longer periods are excited between 6 and 17 months after outburst.  All pulsation periods observed post-outburst are shorter than the dominant quiescent period of 609 s.  These results support the theory that, as the outer convection zone of the white dwarf deepens following the heating of the surface by a dwarf nova outburst, it drives increasing longer periods more efficiently, in proportion to the thermal timescale at the base of the convection zone \citep[e.g.,][]{1983MNRAS.204..537B}. 

Monitoring the pulsation spectrum as V386 Ser quickly cools back towards quiescence enables us to observe a greater range of pulsation periods than are typically excited in a non-interacting pulsating white dwarf, however these may be slightly offset from the quiescent values.  Most notably, the short-period pulsations revealed by the outburst are most sensitive to locations of steep chemical gradients in the core of the white dwarf \citep{2017A&A...598A.109G}, and make this a compelling data set for further asteroseismic analysis.  Another benefit of this system is that the rotational splitting of $1.2\pm0.14\,\mu$Hz observed for the 609\,s mode in V386 Ser \citep{m10} is an order of magnitude smaller than for most pulsating white dwarfs \citep{2017ApJS..232...23H}, resulting in a smaller extrinsic error on the period of the central ($m=0$) component of any multiplets.
Observing which modes are driven to detectable amplitudes as a function of effective temperature following the recent dwarf nova outburst could also be useful for improving the theory of the ZZ Ceti driving mechanism, which is currently unable to predict the energies of individual modes \citep[e.g.,][Section 7.3]{wk08}.

During the interval of the $HST$ observations, several AAVSO members conducted observations spanning several hours.
These datasets were also explored for periodic variability on timescales longer than 5 minutes. However, within the limitations of the smaller telescopes and shorter durations than the McDonald data, no significant periodicities were detected. As GW Lib did not show its 19 min periodicity until about a year past outburst, and it was then intermittently present, we cannot rule out that such a longer mode will appear in V386 Ser. 

\begin{deluxetable}{lccc}[h!]\label{tab:mcdonaldpers}
\tablewidth{0pt}
\tablecaption{Pulsation Periods in UV and Optical Data}
\tablehead{
\colhead{Date} & \colhead{Obs} & \colhead{Period (s)} & \colhead{Amplitude (\%)} } 
\startdata
2019 Jul 03 & opt & 104.00$\pm$0.12 & 0.72$\pm$0.14 \\
2019 Jul 24 & opt & 104.13$\pm$0.14 & 0.63$\pm$0.11 \\
2019 Jul 25 & opt & 104.07$\pm$0.08 & 0.53$\pm$0.08 \\
2019 Aug 15 & UV & 104.284$\pm$0.051 & 2.6  \\
2020 Feb 28 & UV & 174.586$\pm$0.077 & 3.0 \\
2020 Feb 28 & UV & 187.604$\pm$0.090 & 2.0 \\
2020 Jun 22 & opt & 190.0$\pm$0.3 & 0.72$\pm$0.12 \\
\enddata
\end{deluxetable}

\section{Conclusions}

Our ultraviolet spectral observations of V386 Ser at 7 and 13 months after its large amplitude dwarf nova outburst reveal a cooling of the white dwarf by about 2000K during those 6 months. Light curves constructed from the time-tag data show strong periodicity at 104 s in the UV at 7 months post-outburst, and a similar period in the optical data in the month previous to the $HST$ observation. At 13 months post-outburst, the UV data show two longer periods of 174 and 187\,s, and the optical data at 17 months post-outburst varies at 190\,s. Overall, we observe the dominant pulsation periods to increase, appearing to evolve back toward the 609\,s evident during quiescence. This progression from shorter to longer periods follows what is expected from the theory of the cooling of a white dwarf following a dwarf nova outburst. While this is a nice confirmation of pulsation theory, it is too early to tell if the subsequent behavior will continue as a monotonic temperature decrease to the quiescent value of 14,000K, or if V386 Ser will show the unusual cooling over 10 years and the longer period pulsation modes that were evident in GW Lib. Continued observation until the quiescent temperature and pulsation modes are reached is warranted. 

\acknowledgments
PG is pleased to thank William (Bill) P. Blair at the 
Henry Augustus Rowland Department of Physics \& Astronomy at the 
Johns Hopkins University, Baltimore, Maryland, USA, for his indefatigably  
kind hospitality. 
PS and COL acknowledge support from NASA grants HST-GO-15703 and HST-GO-16046 and NSF grant AST-1514737.
KJB is supported by the National Science Foundation under Award AST-1903828.
We acknowledge with thanks the variable star observations from the AAVSO International Database contributed by observers worldwide and used in this research. 
BTG was supported by a Leverhulme Research Fellowship and the UK STFC grant ST/T000406/1. 
OT was supported by a Leverhulme Trust Research Project Grant. 
ZV acknowledges support from NSF grant AST-1707419. PBC acknowledges support from the Wootton Center for Astrophysical Plasma Properties under U.S. Department of Energy cooperative agreement number DE-NA0003843. KIW acknowledges support for McDonald Observatory travel from the NASA K2 Cycle 5 Grant 80NSSC18K0387.
DMT and PK acknowledge support from NASA grants HST-GO-14912 and HST-GO-15316.
This paper includes data taken at The McDonald Observatory of The University of Texas at Austin.
\vspace{5mm}
\facilities{HST(COS), McDonald(ProEM), AAVSO}

\software{IRAF (v2.16.1, \citet{tod93}), 
Tlusty (v203) Synspec (v48) Rotin(v4)  \citep{hub17a,hub17b,hub17c}, 
lmfit (v1.0.1) \citep{lmfit},
PGPLOT (v5.2), Cygwin-X (Cygwin v1.7.16),
W\={o}tan \citep[v1.4;][]{wotan},
MESA \citep[r10398;][]{Paxtonetal2011,Paxtonetal2013,Paxtonetal2015,Paxtonetal2018},
Gyre \citep[v5.2;][]{TownsendTeitler2013},
xmgrace (Grace v2), XV (v3.10) }

\clearpage
\begin{figure}
\vspace{-5cm}
\plotone{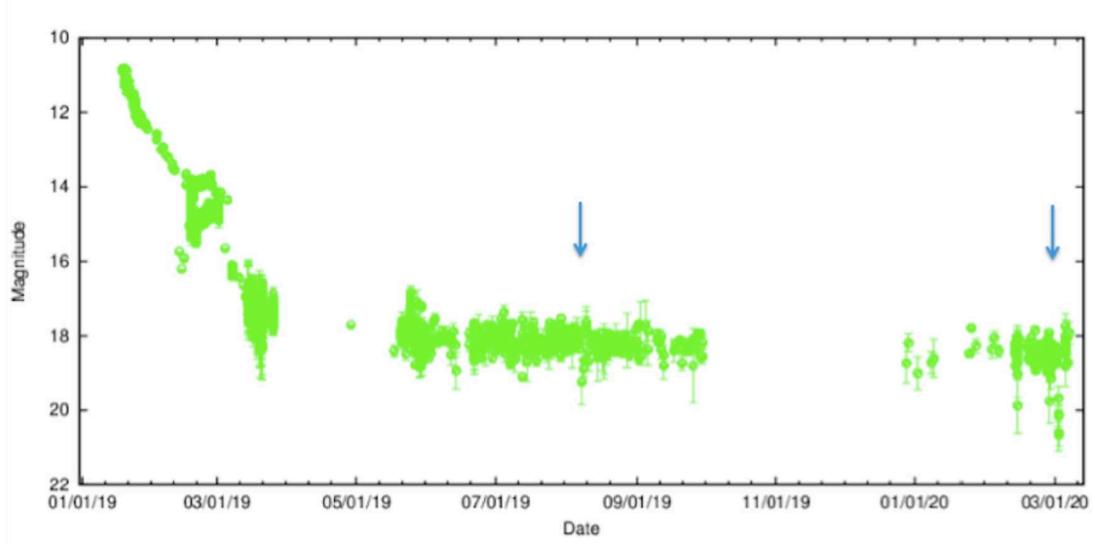}
\vspace{-5cm}
\caption{The AAVSO light curve of V386 Ser. Filled points are V filter, unfilled are clear filter with zeropoint of V filter. The arrows mark the times of the HST observations.}
\label{Figure 1}
\end{figure}

\clearpage
\begin{figure}
\vspace{-11cm}
\plotone{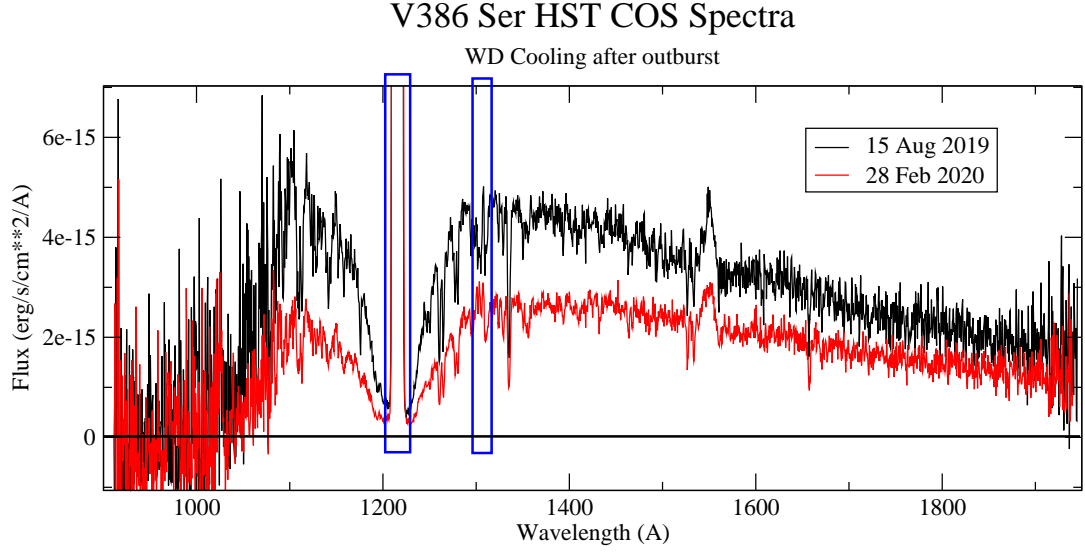}
\caption{The two HST COS spectra of V386 Ser are displayed together for comparison, 
obtained 7 and 13 months after its first known dwarf nova outburst
in Jan 2019. The flux dropped by about half and the 
drop is more
pronounced at short wavelengths, indicating a cooling of the WD temperature. 
The spectra are affected by airglow emission: Ly$\alpha$ ($\sim 1216$ \AA )
and the oxygen doublet ($\sim 1300$ \AA ), both marked in blue.
The C\,{\sc iv} (1550 \AA ) emission is from the source. 
At the edges of the detectors (short and long wavelengths) the signal
is very noisy. Note that the bottom of Ly$\alpha$,
except for the airglow emission, is not at zero, a sign of  
a possible second component.  For clarity, the spectra in the figure 
have been binned at 0.5 \AA , and the region of negative flux ($y < 0$)
is shown.
\label{spectra}
}
\end{figure}

\clearpage 

\begin{figure} 
\vspace{-10.5cm} 
\gridline{\fig{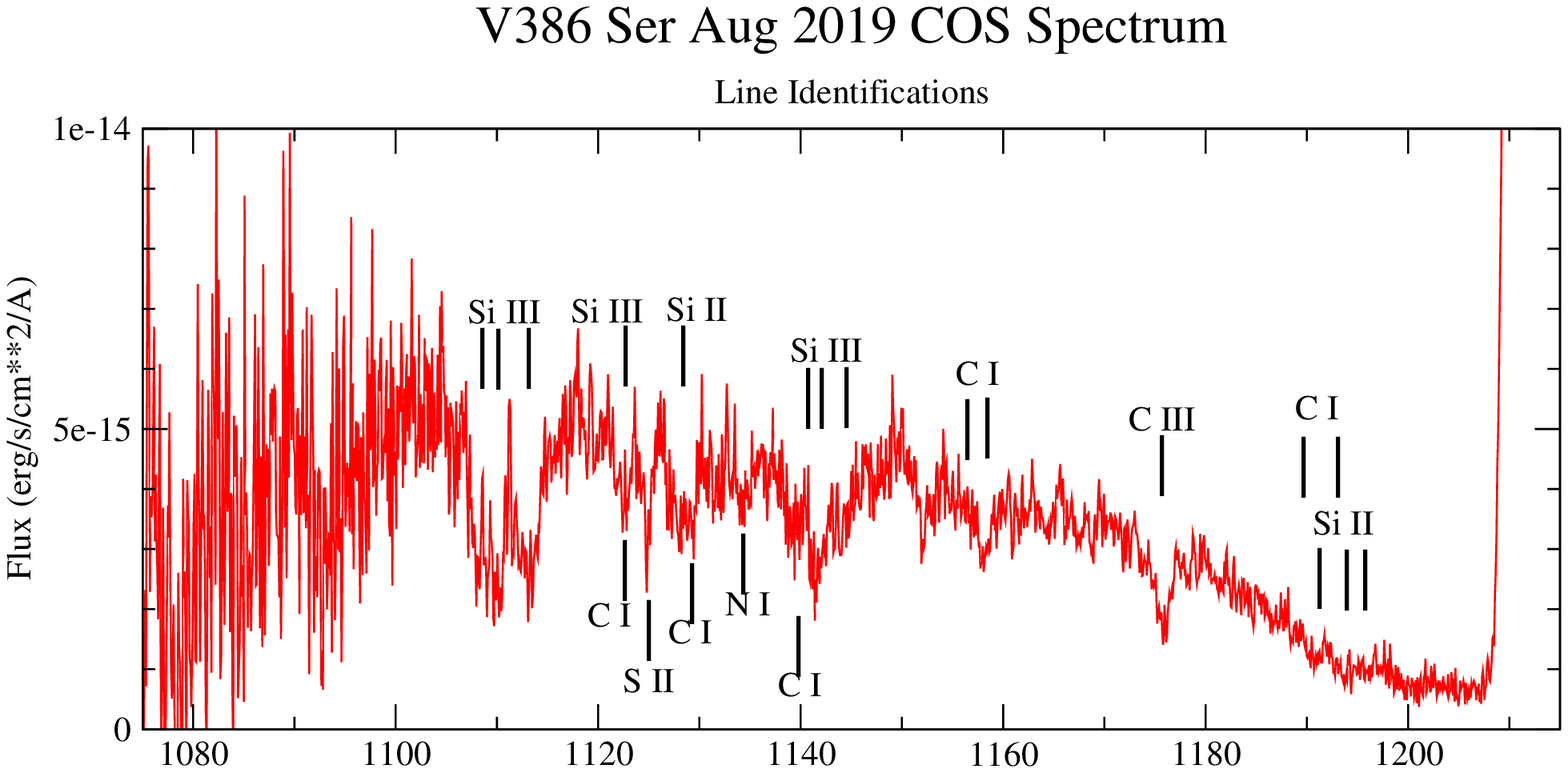}{0.78\textwidth}{}
} 
\vspace{-12.5cm}
\gridline{\fig{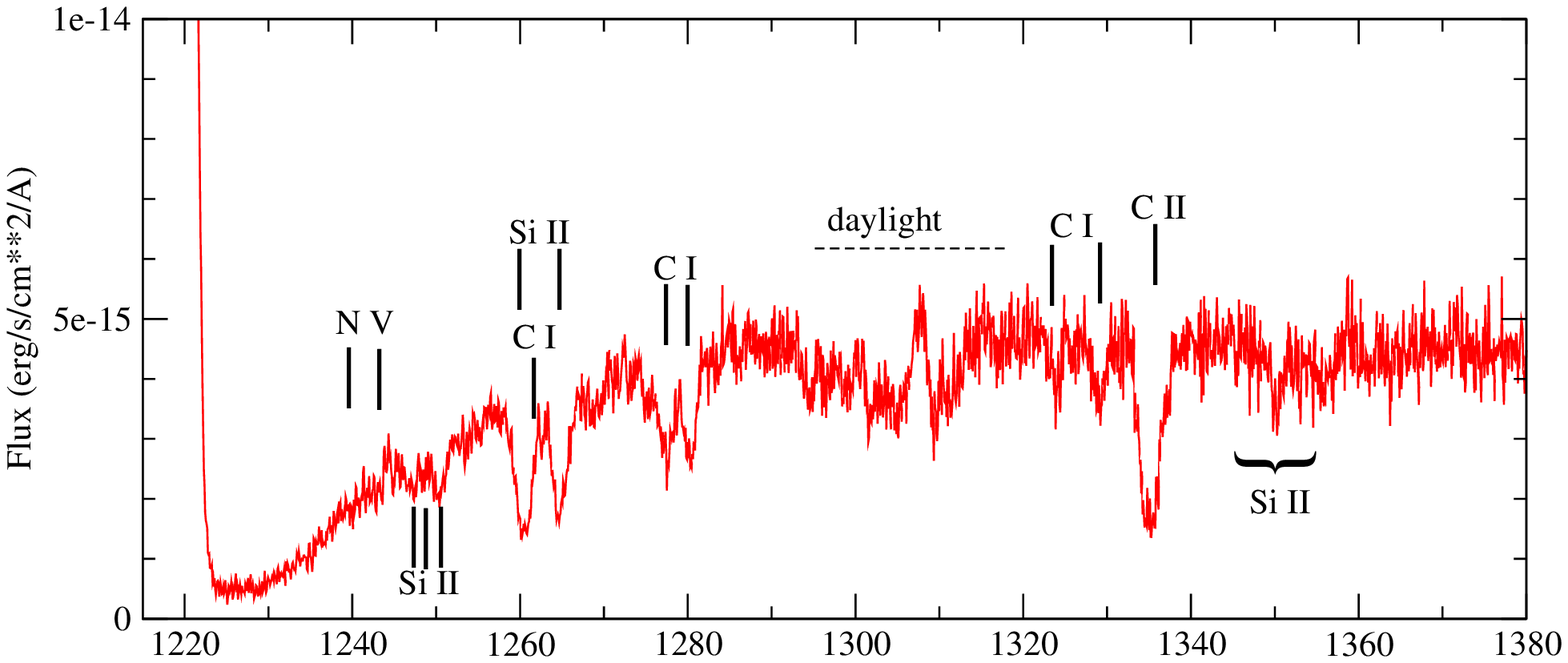}{0.80\textwidth}{}
}
\vspace{-12.5cm}
\gridline{\fig{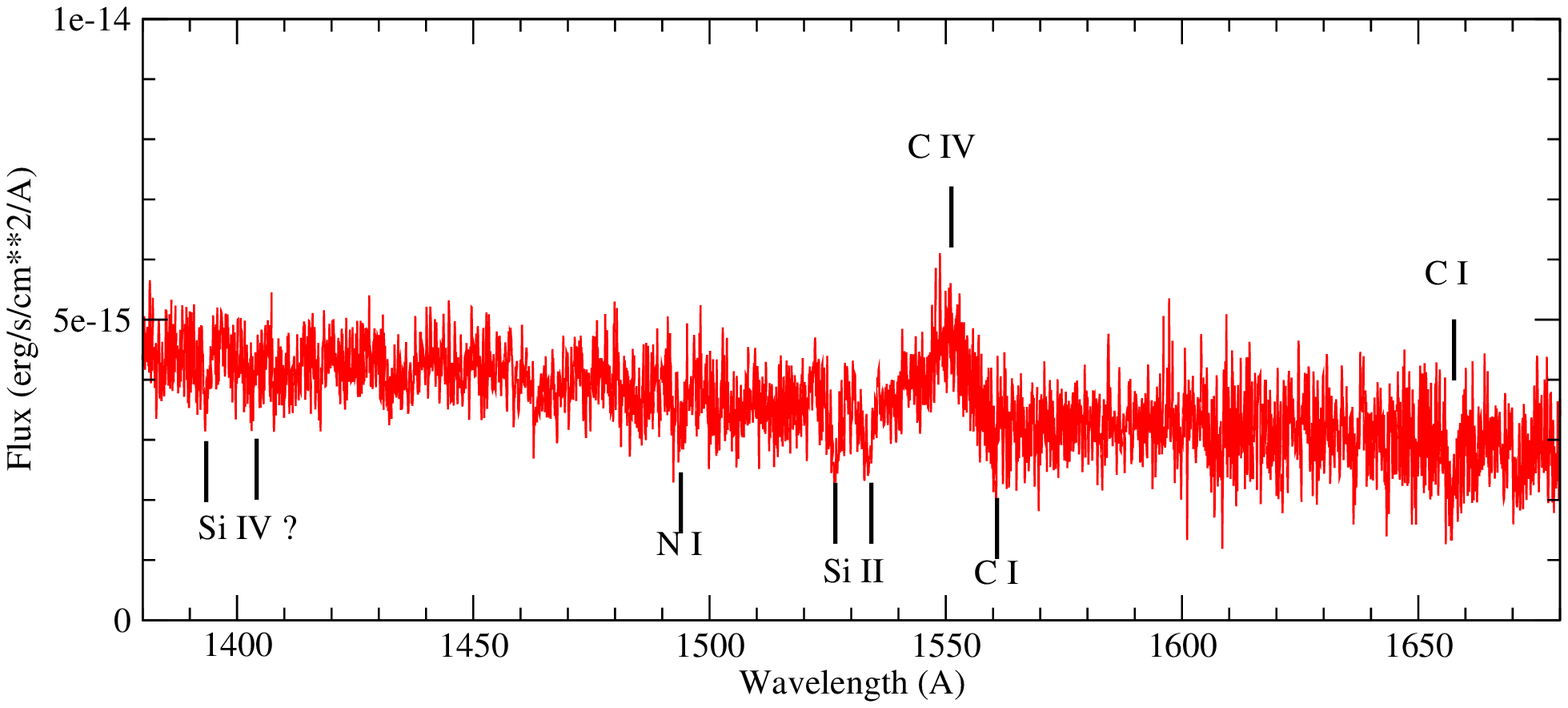}{0.80\textwidth}{}
}
\vspace{-2.cm}
\caption{
We identify the most prominent absorption lines in the 15th August 
COS spectrum of V386 Ser. The same lines are identified in 
the 28 Feb COS spectrum, though not all are as prominent 
as in the Aug 2019 spectrum. 
These absorption lines are used to determine the abundance of elements 
(mainly C \& Si, and tentatively S \& N) and rotational velocity 
of the WD photosphere. Because of the (relatively) low WD temperature,
the Si\,{\sc iv} doublet ($\sim$1400 \AA ) is not observed or is very
weak.   
We do not consider the O\,{\sc i} ($\sim$1300 \AA )
region as it is contaminated with airglow (daylight). We also identify some possible
N\,{\sc v} emission (middle panel).
\label{lineid}
}
\end{figure}

\clearpage 

\begin{figure}
\vspace{-10.cm}
\plotone{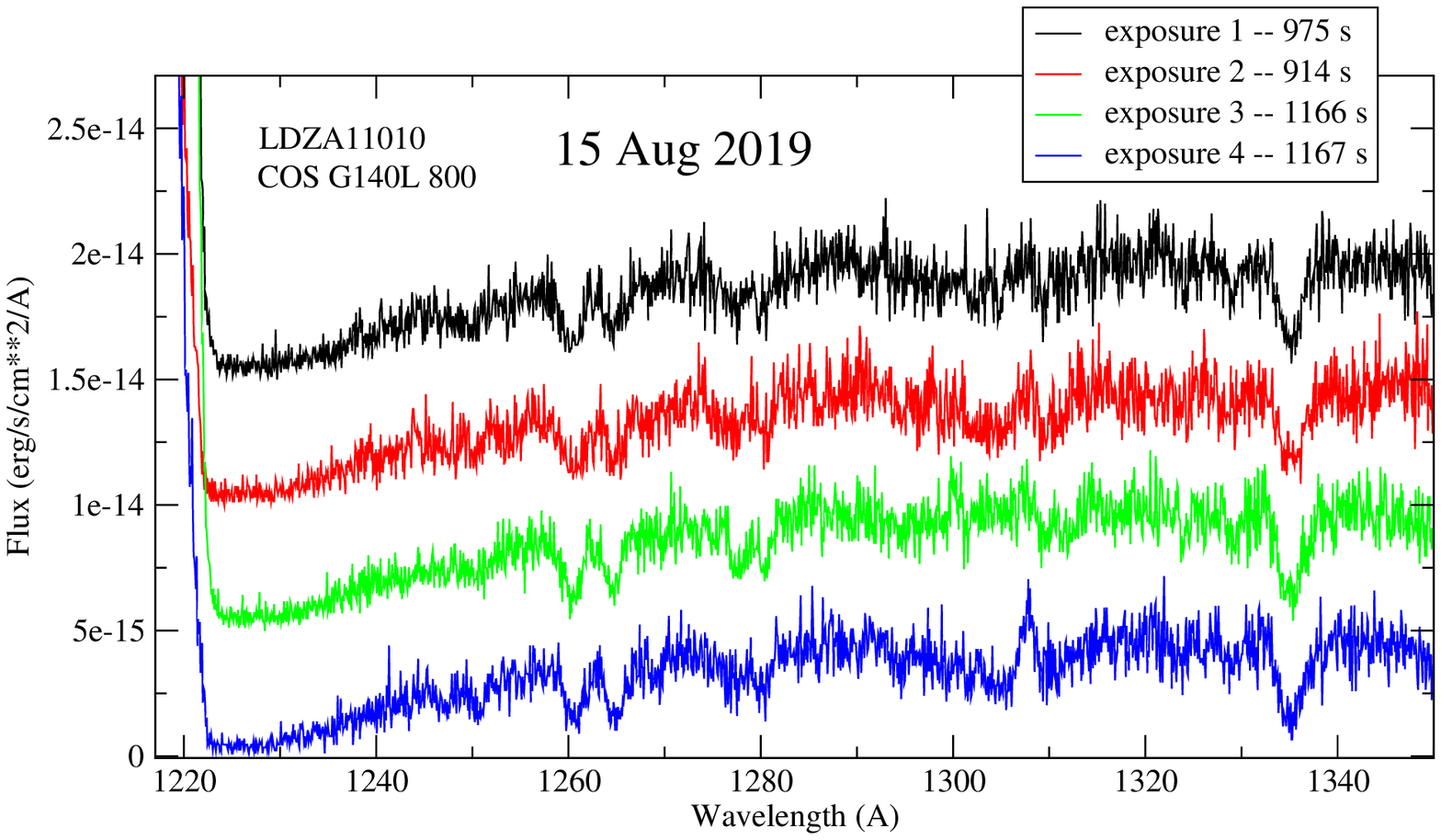}
\caption{The sub-exposures of the two COS spectra 
are shown in the region of the most prominent and 
reliable absorption lines. {\bf Above}: the Aug 2019 spectrum;
{\bf below}: the Feb 2020 spectrum. The spectra have been shifted vertically
for clarity. The good exposure time
is as indicated in the upper right of each panel. 
The shortest exposure time is 310s, 
which corresponds to about 6\% of the binary orbits, 
during which the WD moved $\sim$22$^{\circ}$ 
in its orbital motion. Such a short-time exposure 
is extremely noisy and not reliable to determine 
abundances and the WD rotational velocity. For
that reason that exposure together with the other 
Feb 2020 exposures (for comparison) 
have all been rebinned to 0.2~\AA . 
The original binning is 0.08~\AA\ (as in the upper panel). 
The airglow contamination near 1300~\AA\ is increasing
with the altitude of the sun above the horizon. 
\label{allseg}
}
\vspace{1.cm}
\plotone{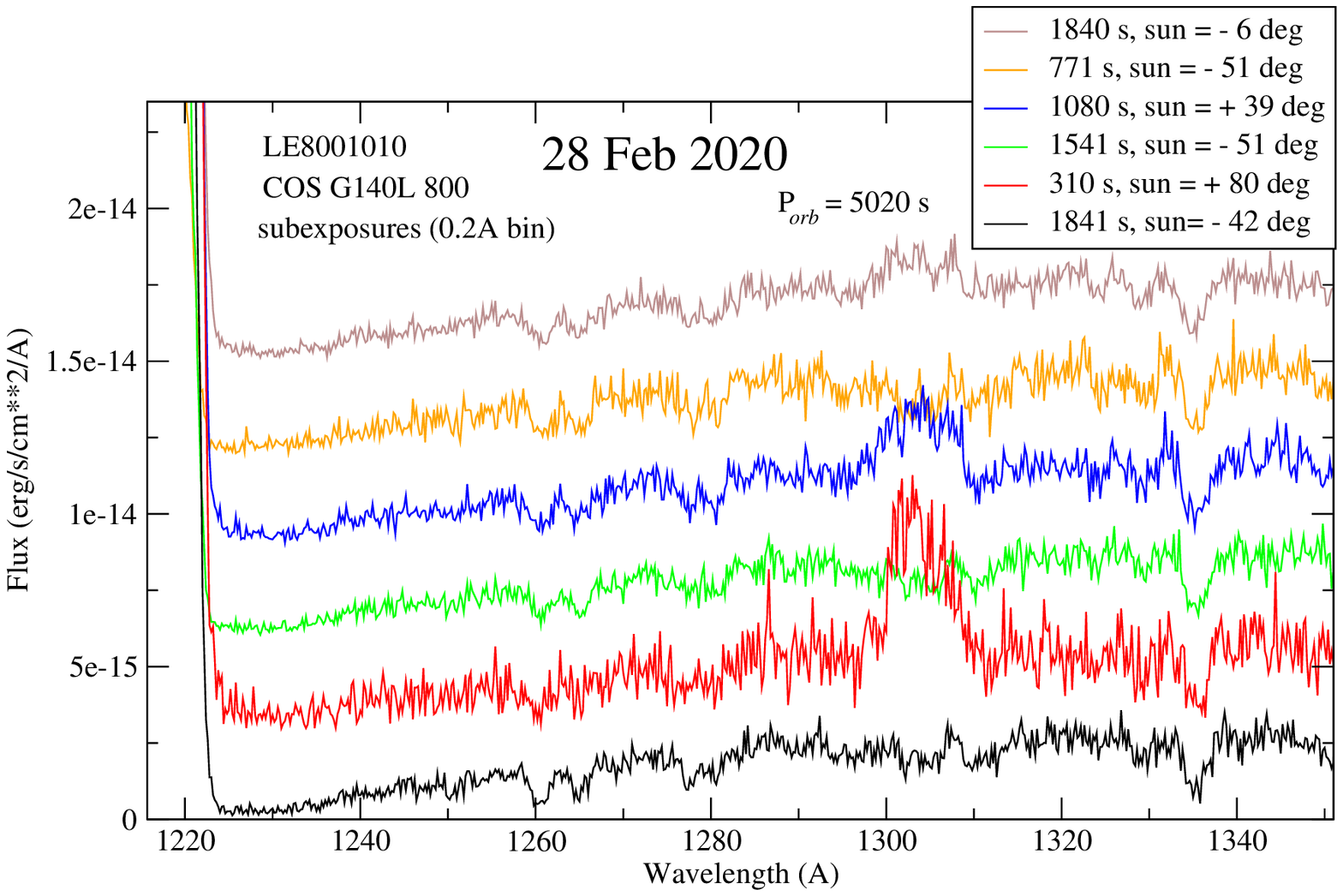} 
\end{figure}

\clearpage
\begin{figure} 
\plotone{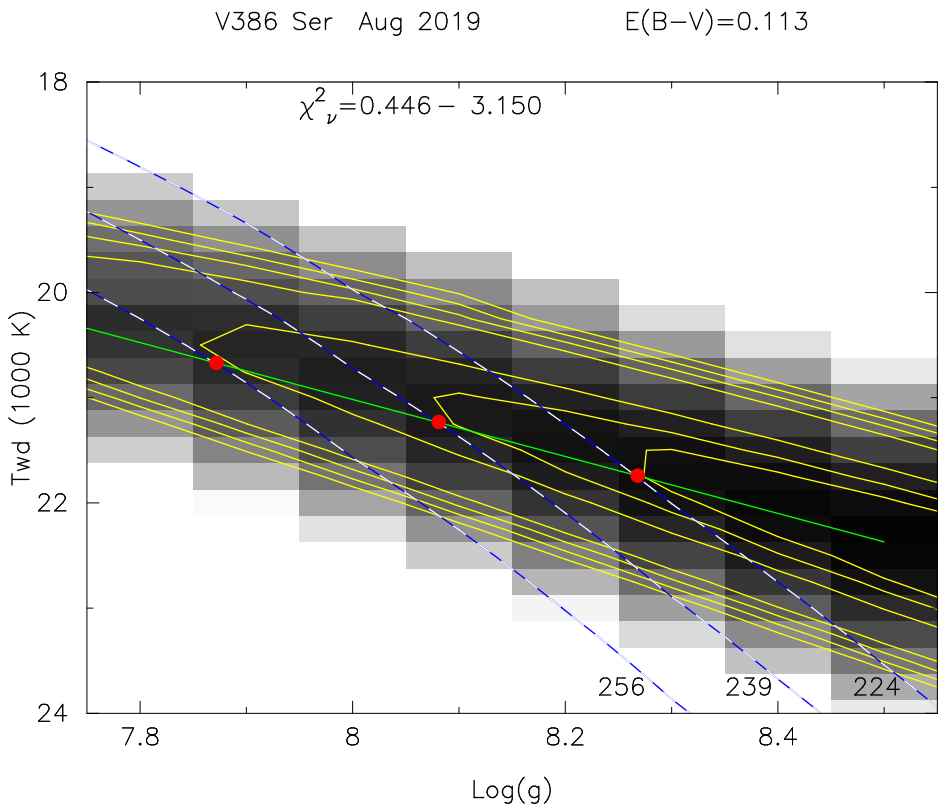}
\caption{
The results of the spectral fit of the Aug 2019 COS spectrum of V386 Ser, 
{\bf dereddened assuming $E(B-V)=0.113$}, are
summarized in this map of the reduced 
$\chi^2_{\nu}$
in the (decreasing) WD effective surface temperature ($T_{\rm wd}$) versus 
(increasing) WD surface gravity ($log(g)$) parameter space.
{\bf The best-fit model for a distance of 239~pc is obtained at the location
of the (middle) red dot on the $d=239$~pc (middle) white blue dashed line 
yielding} $T_{\rm wd}=21,230$~K with  $Log(g)=8.08$. 
{\bf The solutions for a distance of 256~pc and 224~pc are denoted with the
left and right red dots respectively.
The green line connects the best values for the 3 distances considered.
See text for full details.}
\label{aug19achi2} 
} 
\end{figure}

\clearpage
\begin{figure}
\epsscale{1.0}
\plotone{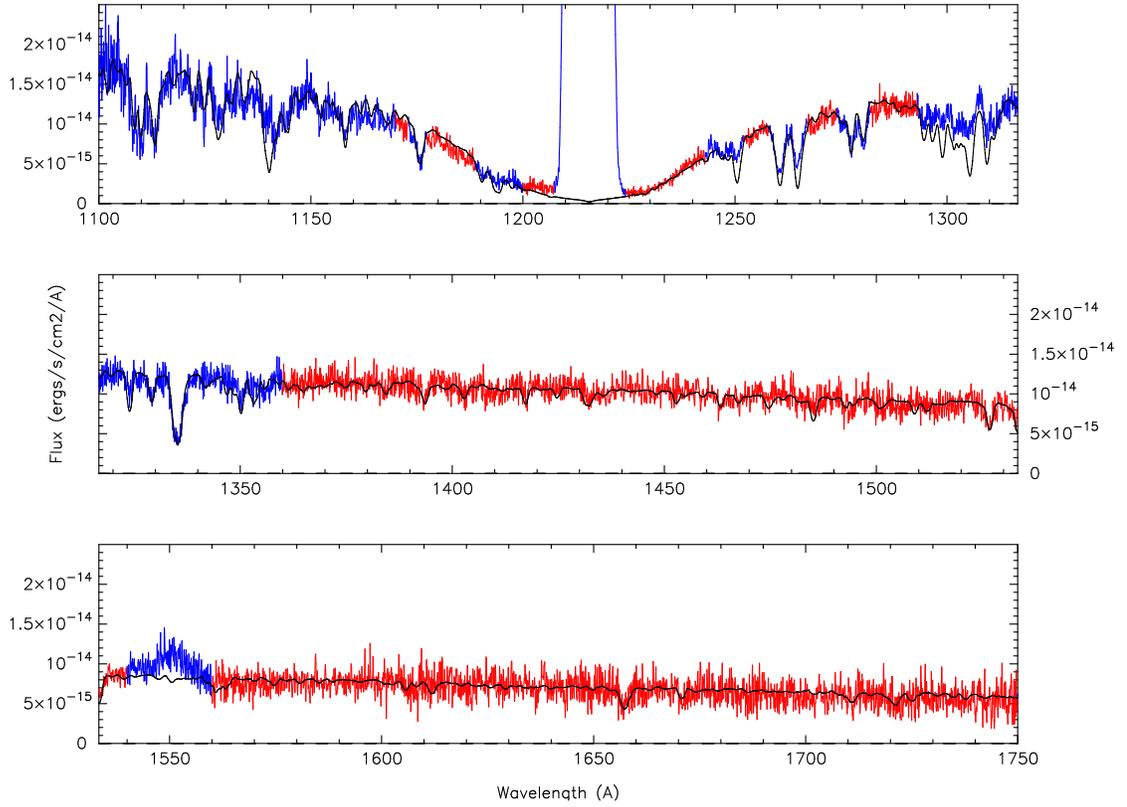}            
\caption{
The Aug 2019 COS spectrum of V386 Ser (red line) fitted with a 
synthetic WD stellar photosphere spectrum (black line). 
The WD  model 
has surface gravity of $Log(g)=8.1$, surface temperature  
of 21,230~K, and scaling to a distance of 239~pc. This model has solar 
composition and a projected rotational velocity of 200 km/s. 
The geocoronal emission regions, the C\,{\sc iv} (1550) emission
line as well as the prominent absorption lines
have been masked before the fitting 
and are marked in blue. The model has excess  flux  
to the left of the Lyman $\alpha$ ($\lambda < 1165$ \AA ). 
This model has $\chi^2_{\nu}=0.475$. 
\label{T21230} 
}
\end{figure}

\clearpage 

\begin{figure}
\plottwo{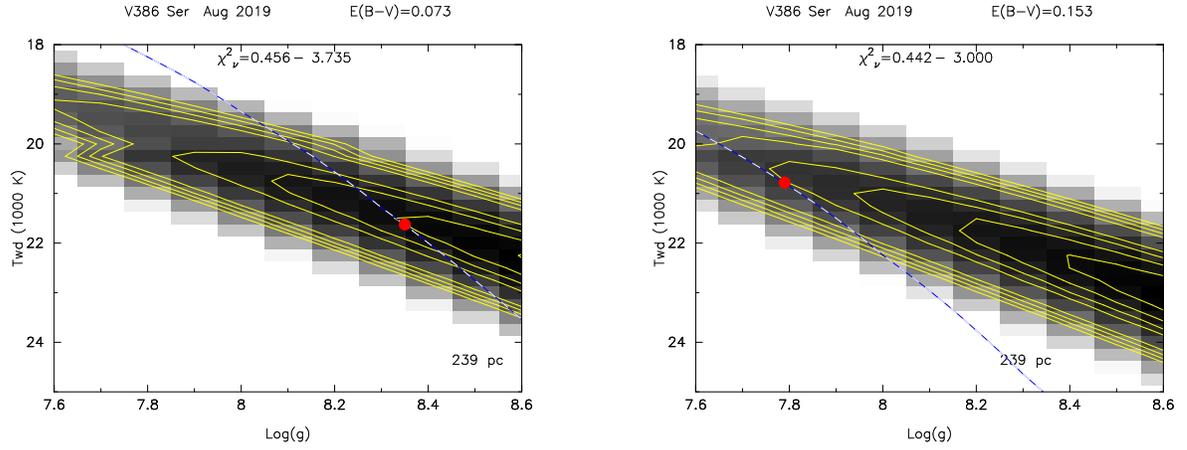}{new_aug19_finegrid_ebv0_153.eps}
\caption{The effect of the propagation of the error in the reddening 
on the solution is illustrated in these two panels. 
The grayscale, contour lines and red dots have the same meaning
as in the previous figure. 
The Aug 2019 spectrum has been dereddened assuming E(B-V)=0.073 
(left panel) and E(B-V)=0.153 (right panel). The overall solution
(dark gray diagonal) has a temperature $\sim 1000$~K higher 
for the larger dereddening. However, as the flux also increases 
for the larger dereddening, scaling to the distance moves toward 
lower gravity, as a larger radius is needed to scale to the 
higher flux at a constant distance. Since the lower gravity 
solutions are colder, the higher dereddening leads to a cooler
WD temperature. 
For E(B-V)=0.073 we obtain a temperature $T_{\rm wd}=21,625$~K 
with $Log(g)=8.35$, while 
for E(B-V)=0.153 we obtain a $T_{\rm wd}=20,780$~K 
with $Log(g)=7.79$. 
\label{aug19_reddening} 
}
\end{figure}

\clearpage
\begin{figure}
\plotone{T21000g8_10Solv200.eps} 
\caption{
As in Fig.\ref{T21230}, 
the Aug 2019 COS spectrum of V386 Ser (red line) is fit with a 
synthetic WD stellar photosphere spectrum (black line). 
A constant flux of $5 \times 10^{-16}$erg/s/cm$^2$/\AA\ has been
subtracted from the spectrum.  
The WD  model has surface gravity of $Log(g)=8.1$, surface temperature  
of 21,000~K, and scaling to a distance of 239~pc. This model has solar 
composition and a projected rotational velocity of 200 km/s.
The excess flux near 1240 \AA\ could be
due to some broad (but weak) N\,{\sc v} emission.   
The model has $\chi^2_{\nu}=0.460$. 
\label{T21000} 
}
\end{figure}

\clearpage

\begin{figure}
\centering
\includegraphics[scale=0.86]{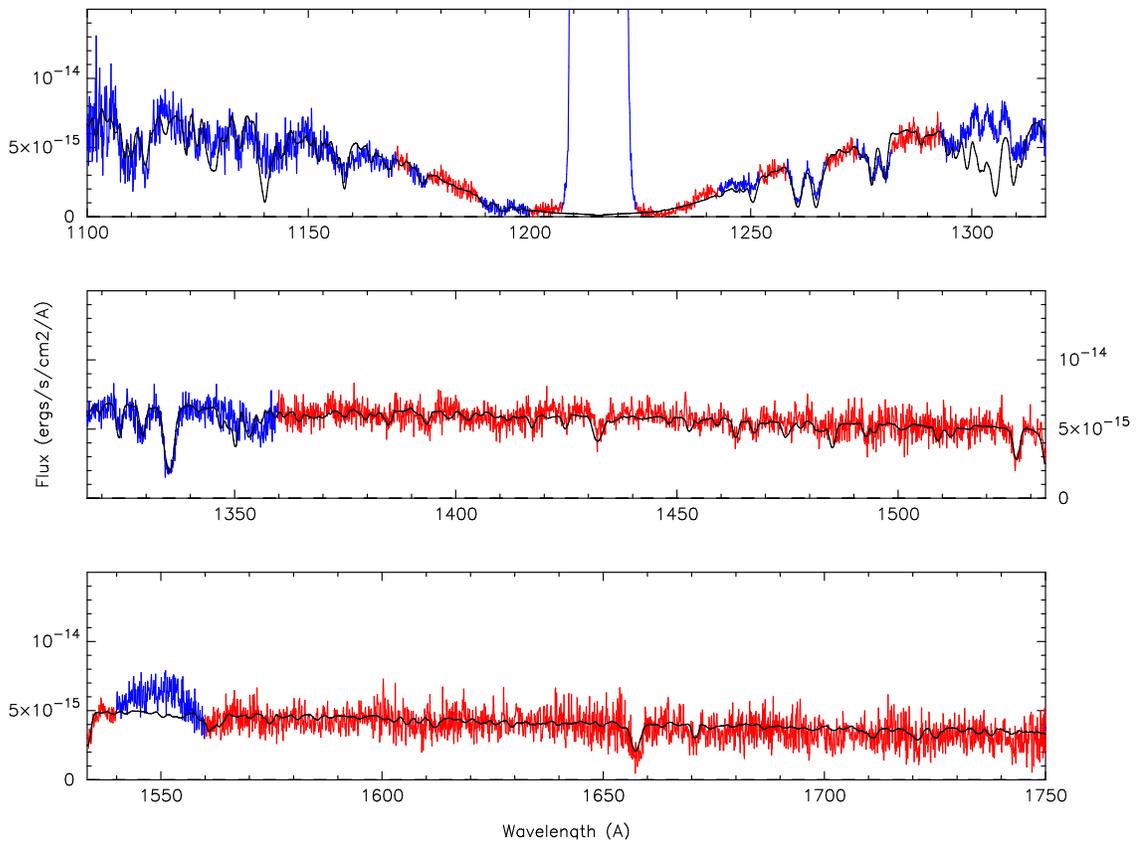}
\caption{
The Feb 2020 COS spectrum of V386 Ser (red line) is fit with a 
synthetic WD stellar photosphere spectrum (black line). 
A constant flux of $6 \times 10^{-16}$erg/s/cm$^2$/\AA\ has been
subtracted from the spectrum.
The WD  model has a surface gravity of $Log(g)=8.1$, surface temperature of 18,750~K, and scales to a distance of 239~pc. 
This model has solar 
composition and a projected rotational velocity of 200 km/s. 
As in the previous figures, the portions of the spectrum in 
blue are the portions that have been masked before the fit.
These are the geocoronal emissions (Ly$\alpha$ and O\,{\sc i} 1300),
and the prominent absorption lines which are being fit separately
from the continuum and Lyman profile.
This model fit has $\chi^2_{\nu}=0.420$. 
\label{T18.75} 
}
\end{figure}

\clearpage

\begin{figure} 
\vspace{-3.cm}
\plotone{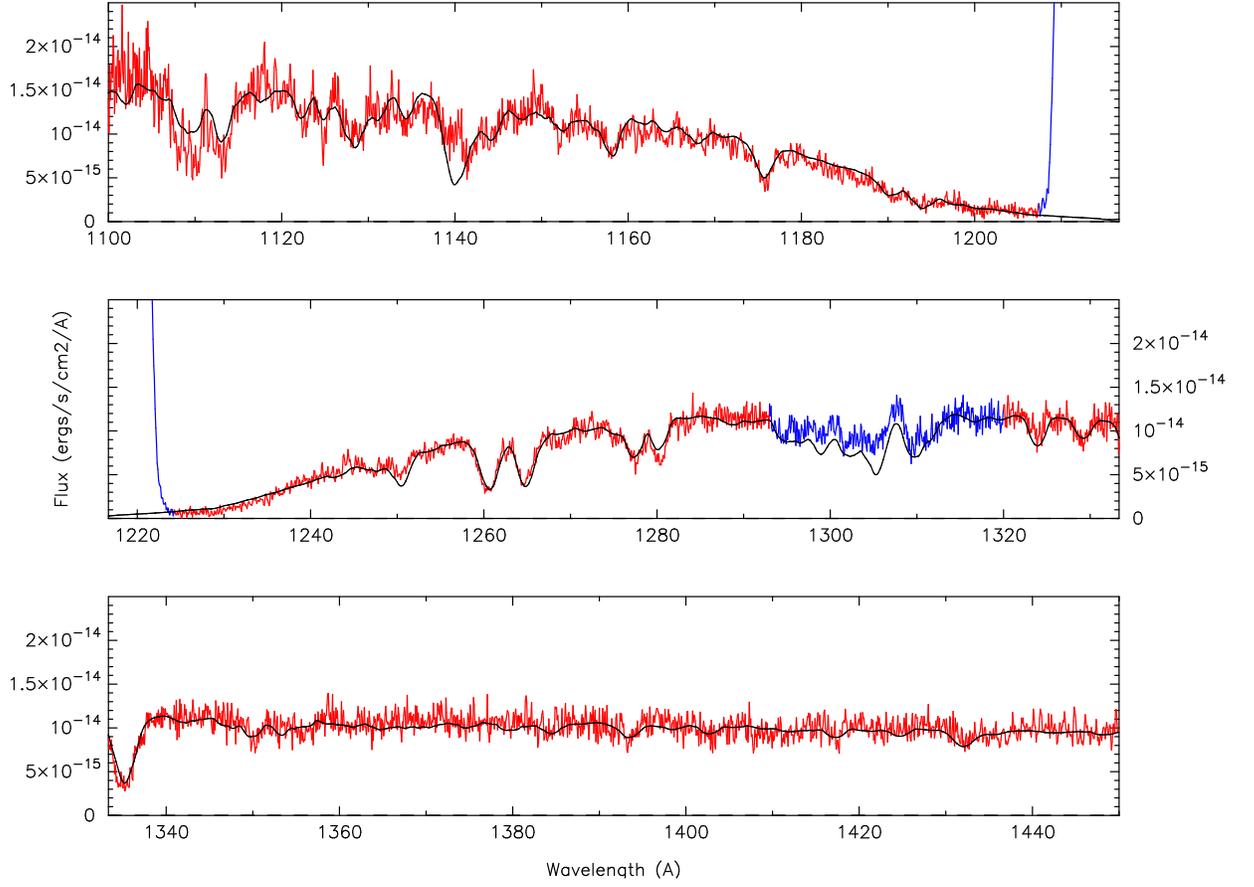} 
\vspace{-1.cm} 
\caption{The Aug 2019 COS spectrum of V386 Ser is modeled with 
a 21,000~K WD with a gravity $Log(g)=8.1$. A constant flux of 
$6 \times 10^{-16}$erg/s/cm$^2$/\AA\ has been subtracted to 
account for a 2nd component. The fitting of the absorption lines
yields a stellar projected rotational (broadening) velocity
of $300 \pm 50$~km/s with solar chemical abundances, except
for silicon [Si]=0.5 (solar). 
\label{aug19ab}
}
\end{figure}

\clearpage
\begin{figure}
\vspace{5.cm}
    \plotone{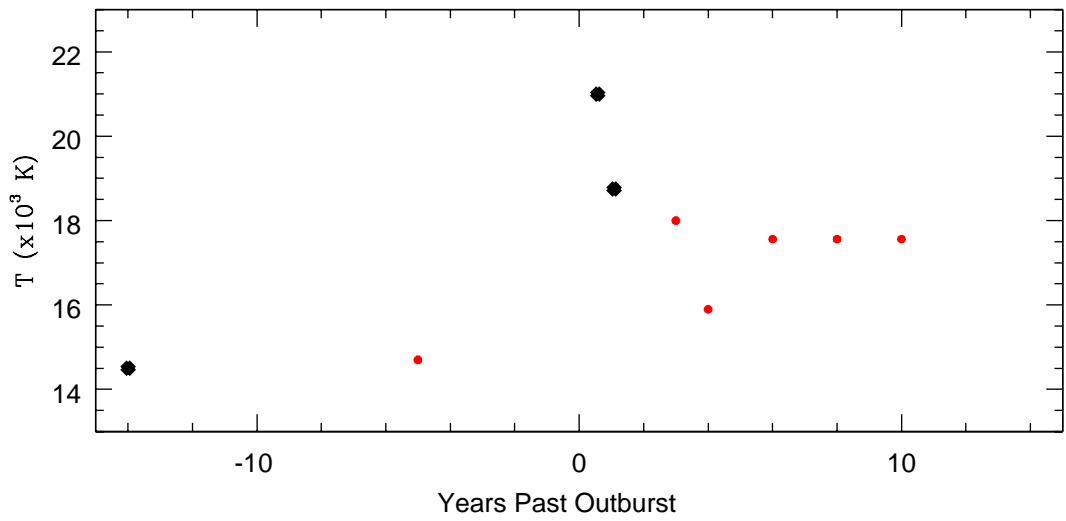}
    \vspace{-5.cm}
    \caption{The WD temperatures determined from UV observations of V386 Ser (large black points) and GW Lib (red points) prior to outburst and in the years after outburst.}
\end{figure}

\clearpage\begin{figure}
    \centering
    \vspace{4.cm} 
    \includegraphics[scale=0.5]{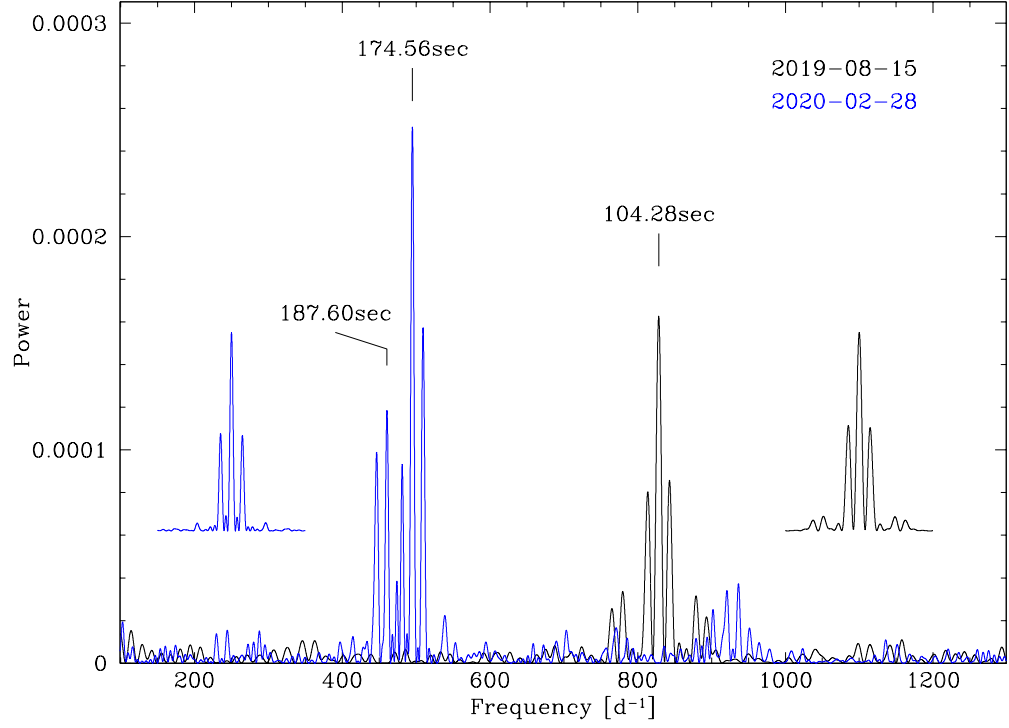}
    \caption{DFT power spectra of time-tag data from August (black) and February (blue) showing the changing pulsation periods between 7 and 13 months after outburst. Insets show the window function for each dataset.}
    \label{fig:powersp}
\end{figure}

\clearpage\begin{figure}
    \centering
    \vspace{4.cm} 
    \includegraphics[scale=0.7]{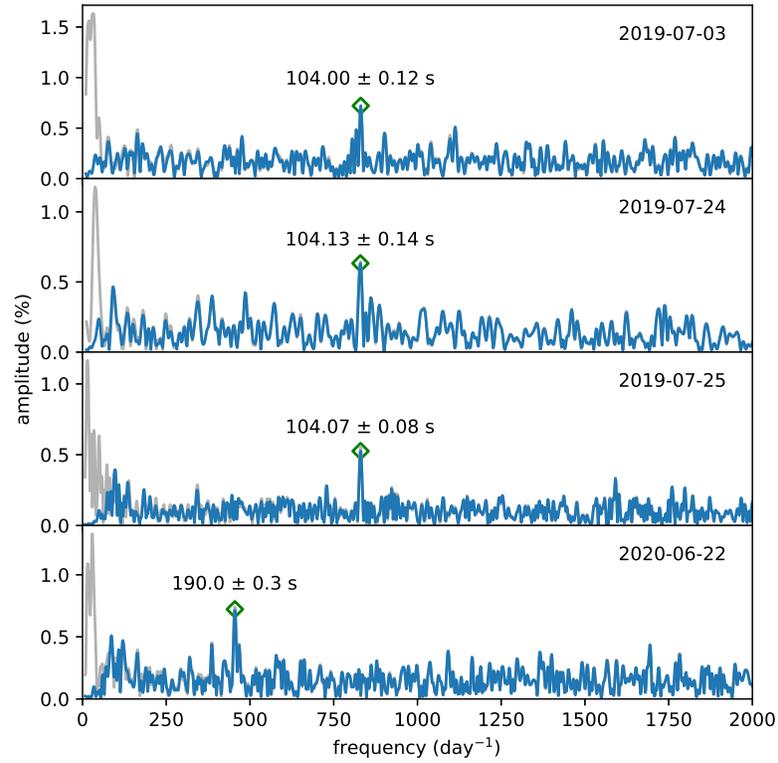}
    \caption{Lomb-Scargle periodograms of each of four nights of time series optical photometry obtained with the ProEM camera on the McDonald Observatory 2.1-meter Otto Struve Telescope through a BG40 filter. Each panel shows the periodogram of the original light curve in light gray, and the darker periodogram after detrending with a 30-minute, 2nd-order Savitzky-Golay filter.  These observations bracket the HST observations and show a similar decrease in dominant pulsation period.}
    \label{fig:mcdonald}
\end{figure}

\clearpage\begin{figure}
    \centering
    \vspace{4.cm} 
    \includegraphics[scale=1.0]{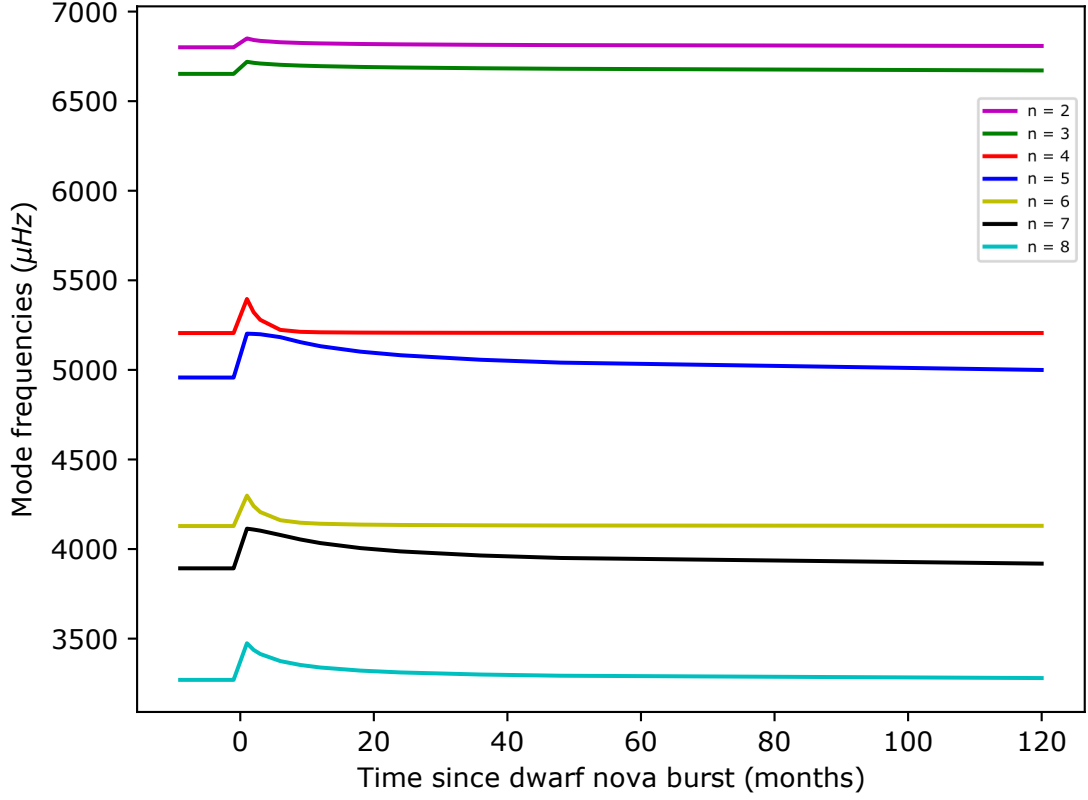}
    \caption{Variations in mode frequencies computed with MESA and GYRE for g-modes of radial order 2-8 in a non-rotating 0.93\,$M_\odot$ white dwarf stellar model in the months following a dwarf nova outburst with a 30 year recurrence time.
    A few percent increment in frequency that slowly relaxes back to the pre-outburst value is expected.}
    \label{fig:mesa}
\end{figure}


\begin{thebibliography}{}

\bibitem[Allard et al.(2020)]{all20} 
Allard, N.F., Kielkopf, J.F., Xu, S. et al. 2020, \mnras, 494, 868

\bibitem[Avni(1976)]{avn76}.
Avni, Y. 1976, \apj, 210, 642 

\bibitem[Arras et al.(2006)]{a06}
Arras, P., Townsley, D. M., Bildsten, L. 2006, \apj, 643, L119

\bibitem[Brickhill(1983)]{1983MNRAS.204..537B} Brickhill, A.~J.\ 1983, \mnras, 204, 537. doi:10.1093/mnras/204.2.537

\bibitem[Capitanio et al.(2017)]{cap17}
Capitanio, L, Lallement, R., Vergely, J.-L. et al.  2017, \aap, 606, 65 

\bibitem[Chote et al.(2021)]{c21}
Chote, P., G\"ansicke, B. T., McCormac, J. et al. 2021, \mnras,tmp..87C

\bibitem[Corsico et al.(2019)]{c19}
Corsico, A. H., Althaus, L. G., Miller, B., Marcelo, M, Kepler, S. O. 2019, AApRev, 27, 7

\bibitem[Debes et al.(2016)]{deb16} 
Debes, J.H., Becker, G., Roman-Duval, J. et al. 2016, 
Instrument Science Report COS 2016-15 (v1) 

\bibitem[Fitzpatrick \& Massa(2007)]{fit07} 
Fitzpatrick, E.L., \& Massa, D. 2007, \apj, 663, 320 

\bibitem[Fontaine \& Brassard(2008)]{fb08}
Fontaine, G. \& Brassard, P. 2008, \pasp, 120, 1043

\bibitem[G\"ansicke et al.(2018)]{gan18}
G\"ansicke, B.T., Koester, D., Farihi, J., Toloza, O. 2018, \mnras, 481, 4323

\bibitem[G\"ansicke et al.(2019)]{g19}
G\"ansicke, B. T., Toloza, O., Hermes, J. J., Szkody, P. 2019, Proc. Conf. Compact White Dwarf Binaries, 2019, eds. Tovmassin, G. H. \& G\"ansicke, B. T. id. 51

\bibitem[Giammichele et al.(2017)]{2017A&A...598A.109G} Giammichele, N., Charpinet, S., Brassard, P., et al.\ 2017, \aap, 598, A109. doi:10.1051/0004-6361/201629935


\bibitem[Godon et al.(2006)]{g06}
Godon, P., Sion, E. M., Cheng, F., Long, K. S., G\"ansicke, B. T., Szkody, P. 2006, \apj, 642, 1018

\bibitem[Godon et al.(2017)]{g17}
Godon, P., Shara, M. M., Sion, E. M., Zurek, D. 2017, \apj, 850, 146
 
\bibitem[Hermes et al.(2017)]{2017ApJS..232...23H} Hermes, J.~J., G{\"a}nsicke, B.~T., Kawaler, S.~D., et al.\ 2017, \apjs, 232, 23. doi:10.3847/1538-4365/aa8bb5

\bibitem[Hippke et al.(2019)]{wotan} Hippke, M., David, T.~J., Mulders, G.~D., et al.\ 2019, \aj, 158, 143. doi:10.3847/1538-3881/ab3984
 
\bibitem[Hubeny (1988)]{hub88}
Hubeny, I. 1988, CoPhC, 52, 103 

\bibitem[Hubeny \& Lanz(1995)]{hub95}
Hubeny, I., \& Lanz, T. 1995, \apj, 439, 875 

\bibitem[Hubeny \& Lanz(2017a)]{hub17a}
Hubeny, I., \& Lanz, T. 2017a, A Brief Introductory Guide to TLUSTY
and SYNSPEC, arXiv:1706.01859 

\bibitem[Hubeny \& Lanz(2017b)]{hub17b}
Hubeny, I., \& Lanz, T. 2017b, TLUSTY User's Guide II: Reference Manual, 
arXiv:1706.01935 

\bibitem[Hubeny \& Lanz(2017c)]{hub17c}
Hubeny, I., \& Lanz, T. 2017c, TLUSTY User's Guide III: Operational Manual,
arXiv:1706.01937 

\bibitem[Kafka(2020)]{K20}
Kafka, S. 2020, Observations from the AAVSO International Database, https://www.aavso.org

\bibitem[Kanaan et al.(2002)]{2002A&A...389..896K} 
Kanaan, A., Kepler, S.~O., \& Winget, D.~E.\ 2002, \aap, 389, 896. doi:10.1051/000/0004-6361:20020485 

\bibitem[Kepler et al.(2000)]{Kepler2000} Kepler, S.~O., Robinson, E.~L., Koester, D., et al.\ 2000, \apj, 539, 379. doi:10.1086/309226

\bibitem[Lampton et al.(1976)]{lam76}
Lampton, M., Margon, B., Bowyer, S. 1976, \apj, 208, 177 

\bibitem[Lindegren et al.(2020)]{lin20} 
Lindergren, L., Klioner, S., Hern\'andez, J., Bombrun, A. et al. 2020, \aap, in prep. 

\bibitem[Luri et al.(2018)]{lur18}
Luri, X., Brown, A.G.A., Sarro, L.M. et al. 2018, \aap, 616, 9 

\bibitem[Mukadam et al.(2010)]{m10}
Mukadam, A. S., Townsley, D. M., G\"ansicke, B. T., Szkody, P., Marsh, T. et al. 2010, \apj, 714, 1702

\bibitem[Newville et al.(2020)]{lmfit}
Newville, M., Otten, R, Nelson, A. et al. 2020, Zenodo, lmfit/lmfit-py 1.0.1 (Version 1.0.1); \url{http://doi.org/10.5281/zenodo.3814709}

\bibitem[Pala et al.(2017)]{pal17}
Pala, A.F., G\"ansicke, B. T., Townsley, D. et al. 2017, \mnras, 466, 2855 

\bibitem[Paxton et al.(2011)]{Paxtonetal2011} Paxton, B., Bildsten, L., Dotter, A., et al.\ 2011, \apjs, 192, 3. doi:10.1088/0067-0049/192/1/3

\bibitem[Paxton et al.(2013)]{Paxtonetal2013} Paxton, B., Cantiello, M., Arras, P., et al.\ 2013, \apjs, 208, 4. doi:10.1088/0067-0049/208/1/4

\bibitem[Paxton et al.(2015)]{Paxtonetal2015} Paxton, B., Marchant, P., Schwab, J., et al.\ 2015, \apjs, 220, 15. doi:10.1088/0067-0049/220/1/15

\bibitem[Paxton et al.(2018)]{Paxtonetal2018} Paxton, B., Schwab, J., Bauer, E.~B., et al.\ 2018, \apjs, 234, 34. doi:10.3847/1538-4365/aaa5a8


\bibitem[Sasseen et al.(2002)]{sas02}
Sasseen, T.P., Hurwitz, M., Dixon, W.V., Airieau, S. 2002, \apj, 566, 267 

\bibitem[Savage \& Mathis(1979)]{sav79}
Savage, B.D., \& Mathis, J.S. 1979, ARA\&A, 17, 73 

\bibitem[Selvelli \& Gilmozzi(2013)]{sel13}
Selvelli, P., \& Gilmozzi, R. 2013, \aap, 560, 49 

\bibitem[Sion(1995)]{s95}
Sion, E. M. 1995, \apj, 438, 876

\bibitem[Sion et al.(1994)]{s94}
Sion, E. M., Lond, K. S., Szkody, P., Huang, M. 1994, \apj, 430, L53

\bibitem[Szkody et al.(2002a)]{sz02a}
Szkody, P., G\"ansicke, B. T., Howell, S. B., Sion, E. M. 2002a, \apj, 575, L79

\bibitem[Szkody et al.(2002b)]{sz02b}
Szkody, P., Anderson, S. F., Ag\"ueros, M., Covarrubias, R. et al. 2002b, \aj, 123, 430

\bibitem[Szkody et al.(2007)]{sz07}
Szkody, P., Mukadam, A., G\"ansicke, B. T., Woudt, P.A., Solheim, J-E. et al. 2007, \apj, 658, 1188

\bibitem[Szkody et al.(2010)]{sz10}
Szkody, P., Mukadam, A., G\"ansicke, B. T., Henden, A., Templeton, M. et al. 2010, \apj, 710, 64 

\bibitem[Szkody et al.(2012a)]{sz12a}
Szkody, P., Mukadam, A. S., G\"ansicke, B. T., Sion, E. M.. Townsley, D. M. et al. 2012a, MmSAI, 83, 513

\bibitem[Szkody et al.(2012b)]{sz12b}
Szkody, P., Mukadam, A. S., G\"ansicke, B. T., Henden, A., Sion, M. et al. 2012b, \apj, 753, 158

\bibitem[Szkody et al.(2016)]{sz16}
Szkody, P., Mukadam, A. S., G\"ansicke, B. T., Chote, P., Nelson, P. et al. 2016, \aj, 152, 48

\bibitem[Timmes et al.(2018)]{Timmesetal2018} Timmes, F.~X., Townsend, R.~H.~D., Bauer, E.~B., et al.\ 2018, \apjl, 867, L30. doi:10.3847/2041-8213/aae70f

\bibitem[Tody(1993)]{tod93} 
Tody, D. 1993, in ASP Conf. Ser. 52, Astronomical Data Analysis
Software and Systems II, ed. R.J. Hanisch, R.J.B. Brissenden, 
\& J. Barnes (San Fransisco, CA;ASP), 173 

\bibitem[Toloza et al.(2016)]{t16}
Toloza, O., G\"ansicke, B. T., Hermes, J. J., Townsley, D. M., Schrebier, M. R. et al. 2016, \mnras, 459, 3929

\bibitem[Townsend \& Teitler(2013)]{TownsendTeitler2013} Townsend, R.~H.~D. \& Teitler, S.~A.\ 2013, \mnras, 435, 3406. doi:10.1093/mnras/stt1533

\bibitem[Townsley at al.(2004)]{t04}
Townsley, D. M., Arras, P., Bildsten, L. 2004, \apj,608, L105

\bibitem[van Zyl et al.(2004)]{vz04}
van Zyl, L., Warner, B., O'Donoghue, D., Hellier, C., Woudt, P. et al. 2004, \mnras, 350, 307

\bibitem[Warner(1995)]{w95}
Warner, B. Cataclysmic Variable Stars, CUP

\bibitem[Warner \& van Zyl(1998)]{wvz98}
Warner, B. \& van Zyl, L. 1998, IAU Symp. 185, 321

\bibitem[Winget \& Kepler(2008)]{wk08}
Winget, D. \& Kepler, S. O. 2008, ARAA, 46, 157

\bibitem[Woods(1995)]{woo95}
Woods, M.A. 1995, in Proc. 9th Europ. Workshop on WDs, 
443, White Dwarfs, ed. D. Koester \& K. Werner (Berlin: Springer), 41 

\bibitem[Woudt \& Warner(2004)]{ww04}
Woudt, P. \& Warner, B. 2004, \mnras, 348, 599




\end{thebibliography}



\end{document}